\documentclass[aps,pre,showpacs,twocolumn,superscriptaddress,floatfix]{revtex4} 
\usepackage{graphicx} 

\begin{document} 
 
\preprint{version 1} 

\title{The electrical double layer for a fully asymmetric electrolyte
around a spherical colloid:\ an integral equation study
} 

\author{G. Iv\'an Guerrero-Garc\'{\i}a}
\affiliation{Instituto de F\'{\i}sica, Universidad Aut\'onoma de San Luis Potos\'{\i}, \\
\'Alvaro Obreg\'on 64, 78000 San Luis Potos\'{\i}, S.L.P., M\'exico} 

\author{Enrique Gonz\'alez-Tovar} 
\affiliation{Instituto de F\'{\i}sica, Universidad Aut\'onoma de San Luis Potos\'{\i}, \\
\'Alvaro Obreg\'on 64, 78000 San Luis Potos\'{\i}, S.L.P., M\'exico} 
\affiliation{Programa de Ingenier\'{\i}a Molecular, Instituto Mexicano del 
Petr\'oleo,\\
Eje Central L\'azaro C\'ardenas 152, 07730 M\'exico, D.F., M\'exico }

\author{Marcelo Lozada-Cassou}
\affiliation{Programa de Ingenier\'{\i}a Molecular, Instituto Mexicano del 
Petr\'oleo,\\
Eje Central L\'azaro C\'ardenas 152, 07730 M\'exico, D.F., M\'exico } 
 
\author{F. de J. Guevara-Rodr\'{\i}guez } 
\affiliation{Programa de Ingenier\'{\i}a Molecular, Instituto Mexicano del 
Petr\'oleo,\\
Eje Central L\'azaro C\'ardenas 152, 07730 M\'exico, D.F., M\'exico } 
 
\date{\today}  
 
\begin{abstract} 
The hypernetted chain/mean spherical approximation (HNC/MSA) integral
equation is obtained and solved numerically for a totally asymmetric
primitive model electrolyte around a spherical macroparticle. The ensuing
radial distribution functions show a very good agreement when compared to
our Monte Carlo and molecular dynamics simulations for spherical geometry and 
with respect to
previous anisotropic reference HNC calculations in the planar limit. We
report an analysis of the potential vs charge relationship, radial
distribution functions, mean electrostatic potential and cumulative reduced 
charge
for representative cases of 1:1 and 2:2 salts with a size asymmetry ratio of
2. Our results are collated with those of the Modified Gouy-Chapman (MGC)
and unequal radius Modified
Gouy-Chapman (URMGC) theories and with those of HNC/MSA in the restricted 
primitive model (RPM)
to assess the importance of size asymmetry effects. One of the most
striking characteristics found is 
that,\textit{ contrary to the general belief}, away
from the point of zero charge the properties of an asymmetric electrical 
double layer (EDL) are not
those corresponding to a symmetric electrolyte with the size and charge of
the counterion, i.e. \textit{counterions do not always dominate}. This
behavior suggests the existence of a new phenomenology in the EDL that
genuinely belongs to a more realistic size-asymmetric model where steric
correlations are taken into account consistently. Such novel
features can not be described by traditional mean field theories like MGC,
URMGC or even by enhanced formalisms, like HNC/MSA, if they are based on the
RPM.
\end{abstract}  
 
  
\maketitle 
  
\section{Introduction}

The electrical double layer (EDL) is the structure formed by electrolyte
ions around a charged surface, usually that of a colloid or electrode. An
understanding of the EDL properties is a crucial matter for science and
technology because of the large variety of related applications, that range
from colloidal stability, electrokinetics and the description of biological
systems to daily manufactured products as inks, paint emulsions, foods or
medicaments \cite{Fe01,Hi01}. As an illustration, the EDL determines 
completely the value
of the zeta potential of a colloid in electrophoretic 
motion \cite{Du01,Hu01}. The zeta
potential, which is directly related to the measured mobility 
in electrophoresis experiments, is a central
quantity in colloid science and many standard techniques of
characterization, stabilization and separation of colloidal suspensions rely
critically on its knowledge \cite{Vo01,Hu02,Ru01,El01}. Depending on the 
form of the dispersed
macroparticle (or electrode) is that we are dealing with an EDL of
particular geometry, e.g. planar, spherical or cylindrical. The planar case
is, by far, the most studied instance, however, the spherical EDL (SEDL)
deserves special attention for its obvious relation to dispersions of
globular proteins, micelles, polymer beads, dendrimers or many other nearly
spherical organic or inorganic macroions \cite{Fe01,Hi01,Vo01,Hu02,Ru01}.

Since the emergence of this topic, the unquestionable relevance of the EDL
has been paralleled by an intense search of an adequate theoretical
description of such charge distribution. In order to build a successful
theory of the EDL, a good model is essential. Among some more refined 
representations,
nowadays, the usual election for the EDL is the minimal but able
restricted primitive model (RPM). In this well-known idealization of a 
\textit{single} EDL a ``colloid'' or ``electrode'' is mimicked by a rigid
and uniformly charged object, which is immersed in an electrolyte solution
constituted by a structureless solvent media and various ionic species,
where the latter are treated as equally-sized hard spheres with punctual
charges at their centers. To avoid image effects it is customary to suppose
a uniform dielectric constant permeating all the space. During the past
decades the RPM-EDL has been studied comprehensively via modern statistical
mechanics approaches, namely integral equations 
\cite{Ca01,Ca02,At01,Lo01,Go01,Vl01,Vl02,Go02,Lo02,Ye01,Yu01,Kj01,Kj02,Pl01,Gr01}, density functional
theories \cite{Mi01,Ki01,Pa01,Pa02,Bo01,Pa03,Bo02,Bo03,Yu02,Wa01}   
and mean electrostatic potential 
schemes \cite{Ou01,Ou02,Ou03,Bh01}. Additionally, the
usage of simulation techniques to evaluate the properties of these coulombic
systems has supplied a large body of ``experimental'' data useful to test the
existing theories 
\cite{To01,To02,Ca02,De01,De02,Te01,Le01,Vl03,Mo01,Ni01,De03}. At the end, 
and on spite of its apparent simplicity, all
the collected evidence sustains the RPM\ as a concise and thriving model
which, by embodying the most important interactions in EDL systems, i.e.
electrostatic and hard-core potentials, is capable to provide not only an
essentially correct account of the associated thermodynamic properties but
also to predict characteristic phenomena of the EDL such as layering, charge
inversion, charge reversal and overcharging. Briefly, we remember to the
reader that the stratified (oscillating) ionic distributions next to a
colloid are referred to as \textit{layering}. On the other hand, 
\textit{charge reversal} means the
excessive compensation of the native colloidal charge, prompted by
strongly attracted counterions, leading to an effective
macroparticle's charge of reversed sign. This counterions layer, in turn, 
promotes a layer of coions adsorption which, consequently, produces 
\textit{charge inversion}, i.e. a layer of inverted charge.
Complementarily, \textit{overcharging}
is the unusual adsorption of coions to the surface of the
colloid, which increases its original charge \cite{Ji01}. In fact, the 
continued progress of theories founded on the RPM has made of them 
a tool of quantitative capacity to interpret experimental results.
As an example, very recently, the new Primitive Model Electrophoresis approach
\cite{Lo03,Lo04} has incorporated consistently the ionic size effects into the
electrokinetic equations and forged an enhanced treatment of
electrophoresis, which fits correctly measurements in multivalent
electrolytic ambients and explains reversed mobilities 
\cite{Lo03,Lo04,Qu01,Qu02}.

Historically, the RPM-based theoretical attempts surged as to better the
classical Poisson-Boltzmann (PB) portrait of the EDL. In the conventional
PB treatment, instead of hard-sphere ions, a punctual electrolyte is
assumed and the ionic distributions are obtained by solving the
Poisson-Boltzmann differential equation, thus eliminating the very important
interionic steric correlations present in the RPM. Although the assumption
of point-ions could be acceptable in some cases (e.g., if low
charged colloids in very diluted univalent salts are being considered), for 
high-coupled systems the inadequacy of such hypotesis has been
clearly attested \cite{To02,Lo01,Go01,Da01,Vl04,Go02}. As a consequence, the 
original \textit{bare} PB\ theory is
unable to predict any of the nonmonotonic characteristics mentioned in the
previous paragraph. Since its early stages, the patent faults of the PB
equation were already noticed \cite{Ki02,Fo01,Re01,Mc01} and, accordingly, a 
first modification to
the simple PB treatment was devised by introducing the concept of the 
Stern layer or the Helmholtz plane. In
this supplemented model, the only inclusion of a \textit{unique} distance of
closest approach between the ions and the macroparticle resulted in the
normalization of the otherwise unphysical values of the radial distribution
functions (RDFs) close to the contact and, eventually, extended the 
applicability of the
PB-Stern or modified Gouy-Chapman theory (PB-S or MGC) to 1:1 
systems with moderate charge or concentration \cite{To03,Ca02}. Nevertheless, 
MGC remained exempt of
nonmonotonical features and, thus, unsuccessful to cope with more demanding
conditions (viz. multivalent and/or highly concentrated electrolytes or,
else, high surface charges).

Up to now, most of the theoretical work on EDL has supposed the valence as the
unique source of ionic asymmetry. Obviously, in nature this is not the
situation and, out of the many possible sophistications to the model (e.g. a
discrete solvent, more species, etc), the consideration of ions with
different sizes is a first choice. Some works have been already published
along these lines \cite{Va01,Bh02,Sp01,Kh01,Ma01,Ou04,Gr02,Va02} but,
notably, the pioneering attempt to include distinct ionic sizes was
formulated inside the framework of the PB equation by Valleau and Torrie 
\cite{Va01},
who used Stern layers of \textit{unequal} extent for the counter- and coions.
It must be noted that in this unequal radius MGC (URMGC) scheme the ions
are, at the same time, voluminous and punctual objects, i.e., an
electrolytic particle behaves as a hard sphere when interacting with the
colloid (being dissimilar the coion and counterion distances of closest
approach to the macroparticle), but among ions they ``see'' each other as
points. In summary, such addition to MGC was rewarding since exposed
interesting effects not observed in the size-symmetric case, e.g. the
displacement of the potential of zero charge (PZC), the occurrence of
oscillations in the radial distribution functions and 
\textit{the apparent dominance of counterions to determine the properties 
of asymmetric 
EDL systems}. At first, all these phenomena seemed appealing since, for 
instance, the non-zero PZC allowed the possibility to interpret experimental 
data by obviating
specific adsorption, and the occurrence of fluctuating URMGC ionic densities
and electrostatic potentials meant that a ``point-ion'' theory should not be
necessarily uniform. Furthermore, for a \textit{n:n} electrolyte with a common 
counterion size, the
convergence between the MGC and URMGC outcomes was so clear that founded the
now widely accepted fact that the counterion size is what really matters for
the EDL. Quoting Valleau and Torrie: ``...\textit{we expect the double layer
properties of a dilute (asymmetric) electrolyte to become similar to those
of a completely symmetric electrolyte having an effective size equal to that
of the counterion. (This remark will be asymptotically exact for large
fields in the Poisson-Boltzmann theory)}...'' \cite{Va01}. Notwithstanding, 
due to the absence of steric interionic
correlations, all the new features in URMGC are constrained in space and can
be rationalized purely in terms of energetic arguments, leaving out
determinant entropic contributions \cite{Lo05,Ji01,Ji02}, so that, 
at the end, URMGC will be
incapable to produce a full description of the phenomenology happening
in general size-asymmetric systems. Even more, we will make evident that 
an analysis  based on the PB viewpoint
of a totally asymetric EDL not only
is partial but, regarding the counterion predominance, can lead in wrong
directions.

For all the stated, to improve the theory of EDL the asymmetry in ionic size
must be included in a more consistent way. That can be accomplished by means
of the \textit{unrestricted} primitive model (PM) of an electrolyte, for
which the condition of equal size and charge in the RPM is relaxed. Not
surprisingly, in the last years, the PM-EDL for \textit{plane} electrodes
has been studied through the avenues of integral equations and the mean
electrostatic potential \cite{Ma01,Ou04,Gr02}. Amid them, the
Greberg and Kjellander's paper \cite{Gr02} represents a valuable contribution 
for its detailed appraisal of the effects of size asymmetry on the ionic
structure and, chiefly, for its thorough discrimination of the different
contributions to charge inversion and to the mean force. In any case, the
body of existent literature on PM-EDL shares two limitations, namely they
have dealt exclusively with the planar geometry, ignoring the 
undoubtfully
notable spherical instance \cite{Tx01}, and have focused on the structural 
properties or
limited themselves to describe semiquantitatively the behavior of the
potential versus charge relationship. Besides, and most
importantly, due to the conditions explored, the majority of 
modern studies going beyond PB \cite{Ca02,Kh01,Ma01,Bo01,Yu02,Va02,Bh03} have
subscribed the wonted conclusion that the counterion-macroparticle
interaction is what determines the EDL properties: ``...\textit{Away from
the pzc, the double layer is dominated by the counterions and, for a binary
system at least, the properties of the double layer are essentially those of
a symmetric salt with the charge and diameter of the counterion}...''
\cite{Kh01}. We will show that this is not always the case.

Consequently, in the present paper we report a comprehensive investigation
of the PM-EDL for a spherical model colloid using integral equations and
simulation techniques with a two-folded aim: i) to produce a
description of the significant PM-SEDL system using a trustable statistical
mechanics treatment and ii) to reivindicate the importance of size asymmetry
in EDL studies in order to get a more faithful picture of a model colloidal
suspension and as the origin of an enriched phenomenology, previously unseen
in the symmetrical conditions. The rest of the paper is organized
as follows. In Section II (and Appendices) we describe the model and give the 
details of the HNC/MSA theory for the PM-SEDL and of the employed simulation 
methods. The results, comparisons and their discussion are included in
Section III, and we close in Section IV with our conclusions and some 
prospectives for future work.
 
\section{Theory and Methods}
\subsection{The Basic Model}

Our representation of the SEDL is constituted by a rigid, charged spherical 
colloid 
of diameter $D$ and surface charge density $\sigma_0$, surrounded by a
continuum solvent of dielectric constant $\epsilon$. The macroion is
in contact with two ionic species which are treated as hard spheres
of {\it diameter} $R_i$ ($i=1,2$) with embedded point charges of 
valence $z_i$ at their centers. \textit{It is assumed for definiteness that 
$R_2 \geq R_1$}. The interaction potential between the macroion and an
ion of type $i$ is then given by

\begin{equation}
\label{umi}
U_{Mi}(r)   = \left\{ \begin{array}{cc}
\infty, & r <  \frac{D+R_i}{2},  \\*[.2cm]
\frac{z_i e 4 \pi \left(\frac{D}{2}\right)^2 \sigma_0}{\epsilon r}, & 
r \geq \frac{D+R_i}{2},
\end{array}
\right. 
\end{equation}

\noindent with $e$ the protonic charge. In turn, the interionic potential is

\begin{equation}
\label{uij}
U_{ij}(r)   = \left\{ \begin{array}{cc}
\infty, & r < \frac{R_i+R_j}{2},  \\*[.2cm]
\frac{z_i z_j e^2}{\epsilon r}, & 
r \geq \frac{R_i+R_j}{2}.
\end{array}
\right. 
\end{equation}

The Stern layer or, more properly, the Helmholtz plane, already mentioned in 
the introduction, is
the geometrical place corresponding to the closest approach
distance for the electrolyte ions to the colloid. If 
we consider an electrolyte formed by a pair of ionic species of 
{\it unequal sizes}, the closest approach 
distance to the surface for the smallest component, $(D+R_1)/2$,
determines the inner Helmholtz plane (IHP), whereas the corresponding 
distance for the largest species, $(D+R_2)/2$, establishes the outer 
Helmholtz plane (OHP). In the limit 
of identical sizes the IHP and OHP coincide and the usual 
definition of the Helmoltz plane is recovered.

\subsection{The HNC/MSA equation for the PM-SEDL}

In general, the Ornstein-Zernike equation and the hypernetted chain closure 
(HNC) for a multicomponent mixture of {\it M} species are

\begin{equation}
\label{ec1}
h_{ij}(r_{12}) = c_{ij}(r_{12}) +
\sum_{l=1}^M \rho_l \int 
h_{il}( r_{13} )c_{lj}( r_{32} ) 
dV
\end{equation}

\noindent and 

\begin{equation}
\label{ec2}
c_{ij}(r_{12}) =  
-\beta U_{ij}(r_{12}) + h_{ij}(r_{12})
- ln ( h_{ij}(r_{12}) + 1 ), 
\end{equation}

\begin{center}
$i,j=1,2 \dots M,$ 
\end{center}

\noindent where $h_{ij}(r_{12})$ is the total correlation function for 
particles 1 and 2 of species $i$ and $j$, distant by $r_{12}$, which is
related to the radial distribution function by 
$g_{ij}(r_{12}) = h_{ij}(r_{12}) + 1$,
$c_{ij}(r_{32})$ is the direct correlation function for the pair 2 and 3,
$\rho_i$ is the bulk number concentration of each one of the species,
$U_{ij}(r_{12})$ is the direct interaction potential, $dV=d^3r_{13}$ is the
differential volume, and $\beta = 1/(k_B T)$,
where $k_B$ is the Boltzmann's constant and {\it T} the absolute temperature.

Applying the Direct Method \cite{Lo06,Lo07,Lo08}, let the species {\it M} 
correspond to macroions at infinite dilution \textit{in a binary 
electrolyte}. Then Eqs. (\ref{ec1}) and (\ref{ec2}) for species 
{\it M} and $j$ become

\begin{equation}
h_{Mj}(r_{12}) = c_{Mj}(r_{12}) +
\sum_{l=1}^2 \rho_l \int 
h_{Ml}( r_{13} )c_{lj}( r_{32} ) 
dV
\label{ec3}
\end{equation}

\noindent and

\begin{equation}
c_{Mj}(r_{12}) =  -\beta U_{Mj}(r_{12}) + h_{Mj}(r_{12}) -
ln ( h_{Mj}(r_{12}) + 1 ), 
\label{ec4}
\end{equation}

\begin{center}
$j=1,2.$ 
\end{center}

\begin{figure}[htbp] 
\begin{center} 
\begin{minipage}{\linewidth} 
{\includegraphics[angle=0.0, width=\linewidth]{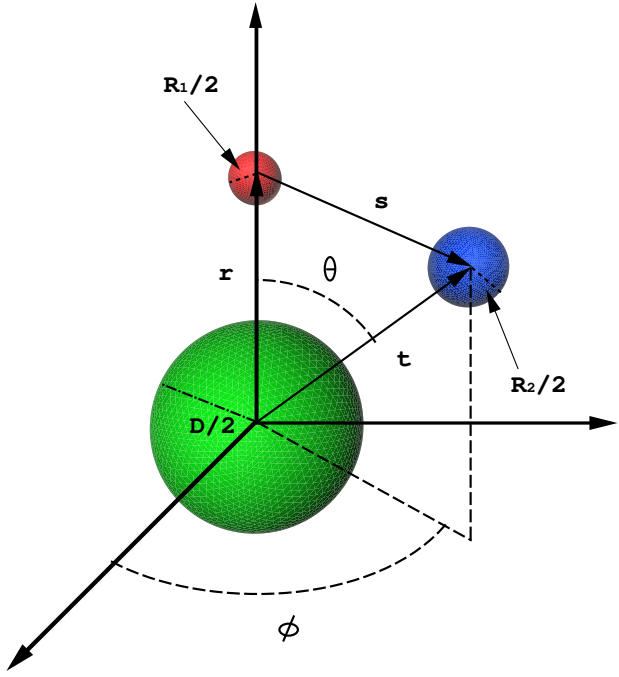}}
\caption{Schematic representation of the model.} 
\label{modelo} 
\end{minipage} 
\end{center} 
\end{figure}

If the HNC closure, Eq. (\ref{ec4}), is used for the macroion-ion direct 
correlation function in the Ornstein-Zernike equation (\ref{ec3}),
we get

\begin{equation}
\label{parch2}
g_{Mj}(r)   = \left\{ \begin{array}{cc}
0, & r < \frac{D+R_j}{2},  \\*[.2cm]
exp \left\{ -\beta U_{Mj}(r) + \Omega_{Mj}(r) \right\}, &
r \geq \frac{D+R_j}{2},  \\*[.2cm]
\end{array}
\right. 
\end{equation}

\noindent with 

\begin{equation}
\label{ome}
\Omega_{Mj}(r) = \sum_{l=1}^2 \rho_l \int \left( g_{Ml}(t) - 1 \right) c_{lj}(s)dV ,
\end{equation}

\noindent where $r=r_{12}$, $t=r_{13}$ and $s=r_{23}=\mid \vec t-\vec r \mid$
are the distances between particles $1$ and $2$, $1$ and $3$,
and $2$ and $3$, respectively, and $dV= t^2~sen\theta~d\theta~d\phi~dt$
is the volume element
in spherical coordinates. The geometry of the system is illustrated
in Fig. \ref{modelo}.

It must be noted that the sizes of the macroion and of the ionic 
species are arbitrary, and 
the only demanded condition is the electrolyte bulk electroneutrality:

\begin{equation}
\label{electroneutralidad}
\rho_1 z_1 + \rho_2 z_2 = 0.
\end{equation}

In respect to the ion-ion direct correlation functions, $c_{ij}(s)$, in 
the bulk MSA theory they can be written as

\begin{equation}
c_{ij}(s) =  c_{ij}^{int}(s) + c_{ij}^{ext}(s). 
\label{cij_msa}
\end{equation}

\noindent These functions are dependent on the size and charge of the ionic 
species and on other parameters of the 
system such as the temperature and the dielectric constant, and are stated 
explicitly in the Appendix A.

Introducing the MSA direct correlation functions in 
Eq. (\ref{ome}), we arrive to   
the following nonlinear system of integral equations (the index {\it M}
is dropped because there is only one macroion present):

\begin{eqnarray}
\label{int1}
g_1(r) & = & exp \left\{ - \beta U_1(r) + 
\rho_1 \int g_1(t) c_{11}(s) dV - \right. {}   
\nonumber \\
& & {} \left. \rho_1 \int c_{11}(s) dV
+ \rho_2 \int g_2(t) c_{21}(s) dV - \right.
\nonumber \\
& & {} \left. \rho_2 \int c_{21}(s) dV \right\}, \;\;\;\; r \geq \frac{D+R_1}{2}, {}
\end{eqnarray}

\noindent and

\begin{eqnarray}
\label{int2}
g_2(r) & = & exp \left\{ - \beta U_2(r) + 
\rho_1 \int g_1(t) c_{12}(s) dV -  \right. {} 
\nonumber \\
& & {} 
\rho_1 \left. \int c_{12}(s) dV + \rho_2 \int g_2(t) c_{22}(s) dV - \right. {}
\nonumber \\
& & {}
\left.
\rho_2 \int c_{22}(s) dV\right\}, \;\;\;\; r \geq \frac{D+R_2}{2}.
\end{eqnarray}

From the specific $c_{ij}(s)$ definition and considering the
electroneutrality restriction,
Eqs. (\ref{int1}) and (\ref{int2}) can be recast in a more compact form as

\begin{eqnarray}
\label{eqsolv1}
g_1(r) & = &exp \left\{ I_0(r) + \rho_1 I_1 (r) - \rho_1 I_2(r) \right. 
{} 
\nonumber \\
{} & &
\left.   + \rho_2 I_3(r) - \rho_2 I_4(r) \right\}
\end{eqnarray}

\noindent and

\begin{eqnarray}
\label{eqsolv2}
g_2(r) & = & exp \left\{ H_0(r) + \rho_1 H_1 (r)  - \rho_1 H_2(r) \right.
{} 
\nonumber \\
{} & & 
\left. + \rho_2 H_3(r) - \rho_2 H_4(r) \right\},
\end{eqnarray}

\noindent with

\begin{equation}
\label{first}
I_0(r)= \frac{-z_1 e 4 \pi {\left(\frac{D}{2}\right)}^2 \sigma_0}{k_B T \epsilon}
\left( \frac{1}{r} \right),
\end{equation}

\begin{equation}
\label{I1}
I_1(r)= \int g_1(t) c_{11}(s) dV,
\end{equation}

\begin{equation}
I_2(r)= \int {c_{11}}^{int}(s) dV,
\end{equation}

\begin{equation}
I_3(r)= \int g_2(t) c_{21}(s) dV,
\end{equation}

\begin{equation}
\label{I4}
I_4(r)= \int {c_{21}}^{int}(s) dV,
\end{equation}

\begin{equation}
H_0(r)= \frac{-z_2 e 4 \pi {\left(\frac{D}{2}\right)}^2 \sigma_0}{k_B T \epsilon}
\left( \frac{1}{r} \right), 
\end{equation}

\begin{equation}
\label{H1}
H_1(r)= \int g_1(t) c_{12}(s) dV,
\end{equation}

\begin{equation}
H_2(r)= \int {c_{12}}^{int}(s) dV,
\end{equation}

\begin{equation}
H_3(r)= \int g_2(t) c_{22}(s) dV,
\end{equation}

\noindent and 

\begin{equation}
\label{H4}
H_4(r)= \int {c_{22}}^{int}(s) dV.
\end{equation}

\noindent The detailed form of these terms is given in the Appendix B.

As a particular case, when $R_1=R_2$ in Eqs. (\ref{umi}) and (\ref{uij}), the 
equations (\ref{eqsolv1}) and (\ref{eqsolv2}) are reduced to 
the HNC/MSA expressions for the RPM-SEDL, deduced by 
Gonz\'alez-Tovar and Lozada-Cassou \cite{Go02}. If the interionic size 
correlations are neglected in the diffuse EDL but 
the closest approach distances between the macroion and the 
ionic species are conserved, by setting  
$R_1=R_2=0$ in Eq. (\ref{uij}) and letting $R_1$ and $R_2$ arbitrary in 
Eq. (\ref{umi}), it is found that
$I_2(r)=I_4(r)=H_2(r)=H_4(r)=0$ and the integral version of the
URMGC theory is obtained. Additionally, if $R_1=R_2=0$ in Eq. (\ref{uij}) and 
$R_1=R_2$ in Eq. (\ref{umi}), the HNC/MSA equations are reduced to the 
corresponding MGC equations in integral form. 

Derived from the ionic profiles, two quantities of special interest are the 
mean electrostatic potential (MEP) and the integrated surface charge density,
which, considering the electroneutrality condition, can be written as: 

\begin{equation}
\psi(r) =   \frac{4 \pi e}{\epsilon} \int_r^\infty \sum_i 
g_i(t) \rho_i z_i \bigg[ t - \frac{t^2}{r} \bigg] dt
\label{pot2}
\end{equation}

\noindent and

\begin{equation}
\sigma(r)= -\frac{e}{r^2} \int_r^\infty \bigg[ \sum_i g_i(t) \rho_i z_i \bigg ] t^2 dt,
\label{qin2}
\end{equation}

\noindent respectively. Such quantities yield important parameters of the 
system when evaluated at specific positions.
For example, if $r=D/2$  Eqs. (\ref{pot2}) and (\ref{qin2}) become
the MEP at the macroion's surface, $\psi_0$, and the native surface charge 
density
of the macroion, $\sigma_0$, respectively:

\begin{equation}
\psi_0 =   \frac{4 \pi e}{\epsilon} \int_{\frac{D}{2}}^\infty \sum_i 
g_i(t) \rho_i z_i \bigg[ t - \frac{t^2}{{\frac{D}{2}}} \bigg] dt
\label{pot_0}
\end{equation}

\noindent and

\begin{equation}
\sigma_0 = -\frac{e}{({\frac{D}{2}}^2)} \int_{\frac{D}{2}}^\infty \bigg[ \sum_i g_i(t) \rho_i z_i \bigg ] t^2 dt.
\label{q_0}
\end{equation}

\noindent Furthermore, for $r=(D+R_1)/2$
Eq. (\ref{pot2}) corresponds to the MEP at the IHP, $\psi_{IHP}$,
whereas for $r=(D+R_2)/2$ the MEP at the OHP, $\psi_{OHP}$, is obtained, i.e.

\begin{equation}
\psi_{IHP} =   \frac{4 \pi e}{\epsilon} \int_{\frac{D+R_1}{2}}^\infty \sum_i 
g_i(t) \rho_i z_i \bigg[ t - \frac{t^2}{{\frac{D+R_1}{2}}} \bigg] dt
\label{pot_ihp}
\end{equation}

\noindent and

\begin{equation}
\psi_{OHP} =   \frac{4 \pi e}{\epsilon} \int_{\frac{D+R_2}{2}}^\infty \sum_i 
g_i(t) \rho_i z_i \bigg[ t - \frac{t^2}{{\frac{D+R_2}{2}}} \bigg] dt.
\label{pot_ohp}
\end{equation}

A fundamental quantity 
in electrokinetics is the zeta potential, $\zeta$. This potential has been
conventionally associated with the MEP at the closest approach
distance between an electrolyte and a charged surface 
\cite{Go01,Go02,Ly01,St01,Ov01,Ly02}. Following
this convention, in a size-asymmetric or PM electrolyte $\zeta=\psi_{IHP}$,
and when the ions are equally-sized 
(in the RPM) $\zeta=\psi_{IHP}=\psi_{OHP}$.

The integrated surface charge density is a very convenient magnitude for 
infinite
surfaces such as unbounded planes or cylinders. Nevertheless, for finite 
surfaces, as spheres, ellipsoids or spherocylinders, is useful to consider 
instead the cumulative reduced charge (CRC):

\begin{eqnarray}
\lefteqn{
Q^*(r)= \frac{Q(r)}{\mid Q_0 \mid} = \frac{1}{4\pi (\frac{D}{2})^2 \mid 
\sigma_0\mid }
 {} } 
\nonumber \\
& & {}
 \Bigg( 4 \pi \bigg ( \frac{D}{2} \bigg ) ^2 \sigma_0 + 4 \pi \int_{\frac{D}{2}}^r \bigg[ \sum_i g_i(t) \rho_i z_i e \bigg ] t^2 dt \Bigg),
{} 
\label{qred}
\end{eqnarray}

\noindent where $Q(r)=4 \pi r^2 \sigma(r)$. Beyond the macroparticle's 
surface this quantity corresponds to the total charge (native plus adsorbed) 
inside a sphere of radius {\it r}, normalized 
with the bare charge over the colloid. If $D/2 \leq r \leq (D+R_1)/2 $
the CRC is equal to one,
whereas for {$r \rightarrow \infty$}  this quantity goes to zero 
because of the electroneutrality condition. The CRC has also the property 
of indicating charge reversal when $Q^*(r)Q_0 < 0$. Moreover, 
charge reversal and overcharging are conveniently quantified by $Q^*(r)$ as a 
fraction or multiple of the
native bare charge compensated locally up to certain distance. 

\subsection{Computational methods}

For a given value of $\sigma _{0}$, the set of nonlinear integral 
Eqs. (\ref{eqsolv1}) and (\ref{eqsolv2}) was numerically solved via the 
successive substitution (or Picard)
method. In the past, this technique has been successfully applied to work out
integral equations associated to similar coulombic fluids 
problems \cite{Lo01,Go01,Yu02}. We have produced accurate results by 
continuing the iterative procedure
until the Euclidean norm between consecutive input and ouput of the RDFs was 
less than some prescribed small number, guaranteeing an
error inferior to 1\%. As an extra test of consistency, at the end of
the Picard cycle, the final ionic profiles were used to check to what extent
they fullfilled the total electroneutrality 
condition in the form of Eq. (\ref{q_0}). In all our calculations 
$\sigma _{0}$ was recovered up to a 99.9 \% at least.

For comparison intentions, the static properties of some of our PM-SEDL
samples were also computed by performing Monte Carlo (MC) and 
molecular dynamics (MD) simulations. In particular, we focused on 2:2, 
0.5~M electrolytes with size asymmetry ratio of 2 
(characterized by the parameters listed in Table \ref{t1}) 
bathing a macroion. This macroion was considered as a hard sphere of
diameter ${\it D}$ = 10 \AA\ with a charge $Z_M e$ located at its center, with 
$Z_M$= 8 and $e$ the protonic charge, which correspond to a surface charge 
density $\sigma _{0}$ = 0.407 C/m$^{2}$. The solvent enters the model as a 
uniform dielectric with an aqueous value of $\epsilon $ = 78.5 at a 
temperature 
${\it T}$ = 298 K. In order to avoid image potential effects the dielectric
constant of the macroion and the ionic species was considered equal
to the solvent. The simulations were carried out within a cubic
box with one macroion fixed in the center, surrounded by the electrolyte. 
The number of ions for each species was adjusted to satisfy
the electroneutrality condition over all system: $N_1Z_1e+N_2Z_2e+Z_Me$=$0$.
The usual periodic boundary conditions with the minimum image convention 
for the ions were imposed to the simulation box, whose length (\textit{L}) was 
considered
sufficiently large to neglect macroion-macroion interactions and to obtain 
stable profiles (the specific parameters used 
are listed in Table \ref{t1}). Following Degr\`{e}ve {\it et al}. 
\cite{De01,De02}, the use of such a sufficiently large
simulation box for SEDL systems owning a plasma parameter 
\cite{Br01,Ca03,Va03,Ro01} 
$\Gamma _{p}<10$ makes unnecessary a treatment 
\`{a} la Ewald of the coulombic interactions. Precisely, in our 
study $\Gamma _{p}=3.1$.

The MC simulations were performed in the canonical ensamble using the 
standard Metropolis method \cite{Al01,Fr01}. The macroion-ion and ion-ion
interaction potentials are given by Eqs. (\ref{umi}) and (\ref{uij}), 
respectively. In the thermalization process were carried
out $4\times10^4$ MC cycles to get the system into equilibrium and 
$2\times10^5$ MC cycles were performed in order to take the canonical 
average after the equilibration. The aceptance ratio used was 0.4. 

In the MD simulations the numerical integration of the motion equations was
accomplished through a reversible MD technique \cite{Tc01}. This method ensures
numerical stability and is equivalent to the velocity Verlet algorithm.
In order to produce data in a feasible way and 
to avoid the problem of an
indefinable force at contact, a very steep but
continuous interparticle potential in lieu of a hard core 
one was employed, i.e.

\begin{equation}
w_{ij}(r)=\left\{
\begin{array}{cc}
A\,\left[ \left( \frac{R_{ij}}{r}\right) ^{2l}-2\left( \frac{R_{ij}}{r}%
\right) ^{l}+1\right] , & r\leq R_{ij}, \\
0, & r>R_{ij},
\end{array}
\right.
\end{equation}

\noindent where $A=k_{B}T$, $l=30$ and $R_{ij}$ is the closest approach 
distance between any pair of particles either macroion-ion or ion-ion. 
In consequence, the total potential among charges is the sum of 
$w_{ij}(r)$ and 
the Coulomb contribution as defined in Eqs. (\ref{umi}) and (\ref{uij}).
The time step used was $\Delta t= 10^{-3}\tau $, where 
$\tau =\sqrt{m_{0}R_{1}^{2}/\left( k_{B}T\right) }$ is the unit of 
time and $m_{0}$ is a characteristic unit of mass. The inertial masses of the 
components of the system are $m_1=m_0$, $m_2= 8.0 \: m_0$ and 
$m_{macroion}= 13.03 \: m_0$. Finally, the formal runs involved more than 
$8\times 10^{6}$ MD steps after equilibration.

\section{RESULTS AND DISCUSSION}

In order to assess the reliability of the HNC/MSA equation for the PM-SEDL
a comparison of our results with simulation data and, possibly, against
alternative theoretical approaches is called for. In the first case,
collating with computer experiments is very useful to discriminate the
adequacy of an approximate theory for a given model system. Additionally,
confronting HNC/MSA with another available and, supposedly, more robust
formalisms allows us to test its degree of accuracy and economy.

Previously, it has been shown that HNC/MSA agrees very well with computer
data for the RPM of planar, spherical and cylindrical EDLs with symmetrical
or asymmerical valences \cite{Lo01,Lo09,Vl01,Vl03,Mo01,De03,De01,De02}. On 
these grounds, now, we have opted for a
certainly strict comparison, namely for a divalent electrolyte with 
appreciable size asymmetry. The two available options for dissymmetry were
considered: counterions smaller than coions and viceversa.

At this point, it is important to mention that even if the equations
developed in the last section are sufficiently general to include asymmetry
in size and charge, here we will restrict our attention to \textit{n:n}
electrolytes (i.e. $\left| \,z_{+}\right| =\left| \,z_{-}\right| $) to
centre on the effect of different sizes. Following this convention,
 \textit{thereinafter, when we refer to PM we will only imply distinct ionic diameters}.

\begin{figure}[htbp]  
\begin{center} 
\begin{minipage}{\linewidth} 
{\includegraphics[angle=0.0, width=\linewidth]{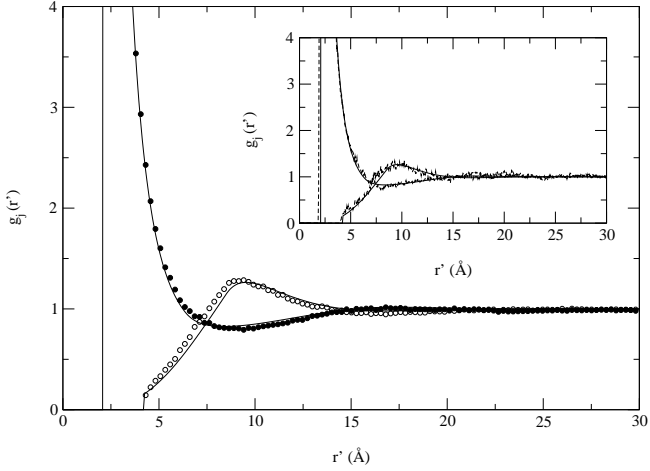}}
\caption{SEDL radial distribution functions for a 2:2, 0.5 M electrolyte 
around a colloid of diameter and surface charge density $D=10$ \AA\ and 
$\sigma_{0}=0.407$ C/m$^{2}$, respectively. The ionic species have
diameters $R_{-}=4.5$ \AA\ and $R_{+}=8.5$~\AA. In the main panel the 
circles and the continuous lines correspond to Monte Carlo data and 
HNC/MSA$_{PM}$ results, respectively. In the inset the same system is 
considered, with
the dashed and continuous lines corresponding to molecular dynamics 
simulations and to the HNC/MSA$_{PM}$ theory, respectively. The 
distance $r^{\prime }$ is measured from the macroparticle's surface.} 
\label{fel1} 
\end{minipage} 
\end{center} 
\end{figure}

\begin{figure}[htbp]  
\begin{center} 
\begin{minipage}{\linewidth} 
{\includegraphics[angle=0.0, width=\linewidth]{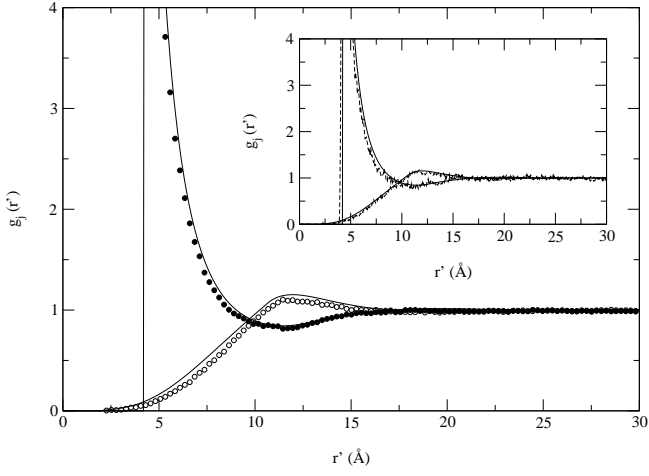}}
\caption{SEDL radial distribution functions for a 2:2, 0.5 M electrolyte 
around a colloid of diameter and surface charge density $D=10$ \AA\ and 
$\sigma_{0}=0.407$ C/m$^{2}$, respectively. The ionic species have
diameters $R_{-}=8.5$ \AA\ and $R_{+}=4.25$~\AA. The symbols and curves
have the same meaning as in Fig. \ref{fel1}.} 
\label{fel2} 
\end{minipage} 
\end{center} 
\end{figure}

In Fig. \ref{fel1} (main panel) we present the radial distribution functions 
for HNC/MSA and MC simulations 
for a macroion of diameter $D=10$ \AA\ and surface charge
density $\sigma _{0}=0.407$ C/m$^{2}$ in a 2:2, 0.5 M electrolyte of
diameters $R_{-}=4.25$~\AA\ and $R_{+}=8.5$ \AA.
The inset of Fig. \ref{fel1} shows RDFs for the same system but
collating HNC/MSA with MD data.
The Fig. \ref{fel2} contains a similar comparison between MC and MD 
simulations and the HNC/MSA theory for the same system considered in the 
Fig. \ref{fel1}, but now for a salt with interchanged ionic
sizes, $R_{-}=8.5$ \AA\ and $R_{+}=4.25$ \AA. From these evidences, it is
rewarding the excellent concordance between theory and simulation for the
two possible asymmetries. 

\begin{figure}[htbp]  
\begin{center} 
\begin{minipage}{\linewidth} 
{\includegraphics[angle=0.0, width=\linewidth]{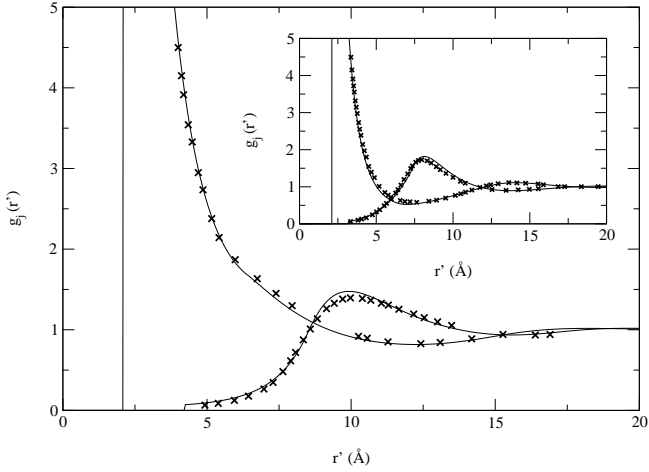}}
\caption{HNC/MSA$_{PM}$ and ARHNC radial distribution functions
for monovalent and divalent 1 M electrolytes. For HNC/MSA$_{PM}$ 
(continuous lines) the ionic
distributions are around a very large colloid of diameter $D=1000\times R_{-}
$ and surface charge density $\sigma _{0}=0.267$ C/m$^{2}$. For the
ARHNC theory (with symbols) the profiles are next to a wall of
the same charge $\sigma _{0}=0.267$ C/m$^{2}$. In the main panel the 1:1 ionic
species have diameters $R_{-}=4.25$ \AA\ and $R_{+}=8.5$ \AA\, and in the
inset the 2:2 ionic diameters are $R_{-}=4.25$ \AA\ and $R_{+}=6.375$~\AA. 
The distance $r^{\prime }$ is measured from the charged surface.} 
\label{in_kjell} 
\end{minipage} 
\end{center} 
\end{figure}

Complementarily, we also made some comparisons between the spherical HNC/MSA
in the limit $D\rightarrow \infty $ and published data of the anisotropic 
reference HNC theory (ARHNC) for
a charged planar wall \cite{Gr02}. We resorted to this limit because, to the 
best of our knowledge, there are no alternate structural results for a 
charged macrosphere
in a size-asymmetrical electrolyte \cite{Kh01}. Furthermore, at least for the 
planar
geometry, ARHNC is an accurate description that considers the inhomogeneous
correlations in the electrolyte due to the wall and that fits very closely
the simulations of the RPM-EDL. Thus, in Fig. \ref{in_kjell} (main panel) 
are displayed the HNC/MSA
ionic profiles for a huge macroion of diameter $D=1000\times R_{-}$ and
charge density $\sigma _{0}=0.267$ C/m$^{2}$, surrounded by a 1:1, 1 M
electrolyte with $R_{-}=4.25$~\AA\ and $R_{+}=8.5$ \AA, and the ARHNC RDFs 
corresponding to a charged wall bearing the same $\sigma _{0}$ and under
identical electrolytic parameters. To cover a more exigent situation, in
the inset of Fig. \ref{in_kjell} we plot the HNC/MSA distributions of a 
2:2, 1 M
electrolyte of ionic dimensions $R_{-}=4.25$~\AA\ and $R_{+}=6.375$ \AA\
around a large colloid, characterized by $D=1000\times R_{-}$ and $\sigma
_{0}=0.267$ C/m$^{2}$, along with the ARHNC outcomes for the same electrolyte
in contact with a wall of equal $\sigma _{0}$. Clearly, for both the uni-
and divalent systems, the coincidence between HNC/MSA\ and ARHNC is notable,
considering that HNC/MSA is a simpler and less computationally demanding 
theory.
Summarizing, the comparisons with Monte Carlo, molecular dynamics and ARHNC 
prove the
trustability of HNC/MSA for the PM-SEDL and gives us the confidence to
explore its predictions in a wider range of conditions.

A global and concise manner to analize the properties of the EDL for a
large set of different states is in terms of the relationship between the MEP
at some some point and the surface
charge. In this respect, the first charge derivative of the MEP at the 
surface, $\frac{d\psi _{0}}{d\sigma _{0}}$, is relevant because of its
connection with the differential capacity 
\cite{Bo03,At02,To04,Pa04,Bo04,Ea01,Go03},
whereas the MEP at the IHP, $\psi _{IHP}$, is conventionally associated with
the zeta potential, $\zeta $, a key quantity in electrokinetic
phenomena. Besides, and as it will be shown, it is precisely in the behavior
of these MEPs that the importance and extension of the size asymmetry effects 
under
study are evidenced. Because of this, we have calculated the functions $\psi
_{0}(\sigma _{0})$ and $\psi _{IHP}(\sigma _{0})$ for a size-asymmetric
SEDL using the following four theoretical schemes: URMGC, MGC, HNC/MSA in
the PM (HNC/MSA$_{PM}$), and HNC/MSA in the RPM (HNC/MSA$_{RPM}$). In a
series of preliminary runnings we have noticed that the manifestations of
size asymmetry are more emphatic when the counterions are smaller than
coions, thus, below \textit{we shall concentrate on the case where the
diameter of the coions is twice that of the counterions}. For comparison
purposes, the parameters of the systems are adjusted such as for URMGC and
HNC/MSA$_{PM}$\ the unequal colloid-counterion and colloid-coion closest
approach distances are shared, and for MGC and HNC/MSA$_{RPM}$ the unique
closest approach distance corresponds to that employed in HNC/MSA$_{PM}$ 
\textit{for colloid-counterion}. Therefore, in the rest of our calculations
when the ionic diameters of a PM electrolyte, $R_{-}$ and $R_{+}$, are
stated it is meant that the different theories are solved for the following
elections of the colloid-ion and ion-ion distances of closest approach, $%
d_{i}$ and $d_{ij}$, respectively (with $\sigma _{0}\geq 0$):

\begin{equation}
d_{i}=\left\{ 
\begin{array}{ll}
d_{-}=\frac{D+R_{-}}{2}~ \text{and}  ~d_{+}=\frac{D+R_{+}}{2}\text{,} & 
 \text{for HNC/MSA}_{PM} \\
& \text{ and URMGC,} \\ 
d_{-}=d_{+}=\frac{D+R_{-}}{2}\text{,} & \text{for HNC/MSA}_{RPM} \\
&\text{and MGC,}
\end{array}
\right.
\end{equation}

\begin{equation}
d_{ij}=\left\{ 
\begin{array}{ll}
d_{-\,+}=d_{+\,-}=\frac{R_{-}+R_{+}}{2}\text{,} & \text{for HNC/MSA}_{PM}
\text{,} \\ 
d_{-\,+}=d_{+\,-}=R_{-}\text{,} & \text{for HNC/MSA}_{RPM}\text{,} \\ 
d_{-\,+}=d_{+\,-}=0\text{,} & \text{for URMGC and MGC.}
\end{array}
\right. 
\end{equation}

\begin{figure}[htbp] 
\begin{center} 
\begin{minipage}{\linewidth} 
{\includegraphics[angle=0.0, width=\linewidth]{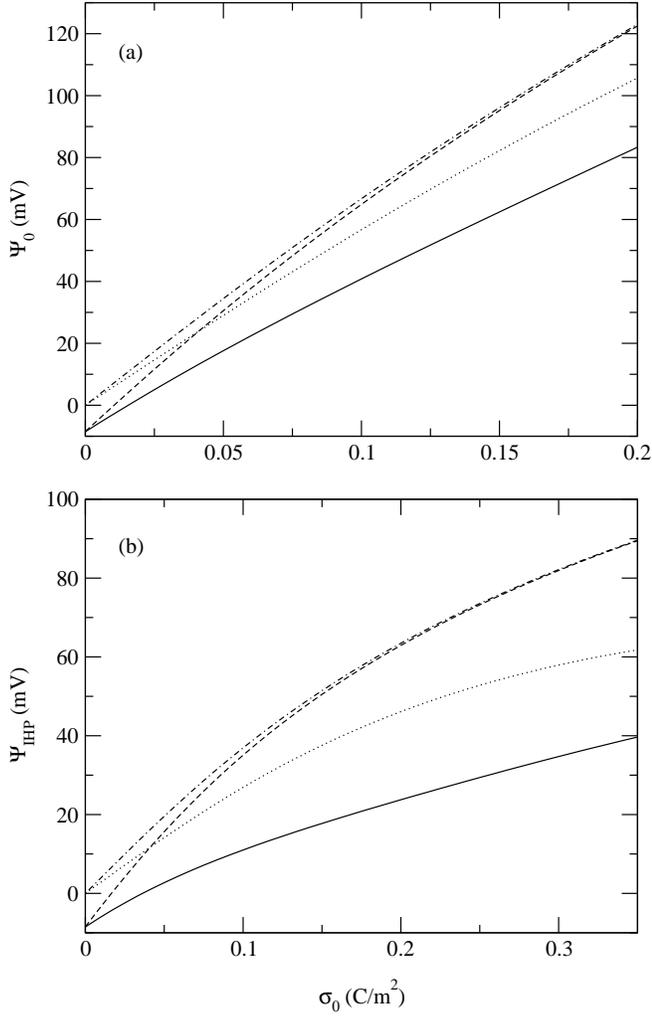}}
\caption{SEDL mean electrostatic potential at: [a] the surface of the colloid
and [b] the inner Helmholtz plane, as a function of the surface charge
density $\sigma _{0}$, for a 1:1, 1 M electrolyte with $R_{-}=4.25$~\AA\ and
$R_{+}=8.5$~\AA\ around a macroion of diameter $D=160$ \AA\ and non-negative
surface charge density. The solid and dotted lines stand for HNC/MSA$_{PM}$
and HNC/MSA$_{RPM}$, and the dashed and dot-dashed lines correspond to URMGC
and MGC, respectively. The ion-colloid closest approach distances are: 
$d_{-}=(D+R_{-})/2$ and $d_{+}=(D+R_{+})/2$ for HNC/MSA$_{PM}$\ and URMGC,
and $d_{-}=d_{+}=(D+R_{-})/2$ for HNC/MSA$_{RPM}$ and MGC. The ion-ion
closest approach distances are: $d_{-\,+}=d_{+\,-}=(R_{-}+R_{+})/2$ for
HNC/MSA$_{PM}$, $d_{-\,+}=d_{+\,-}=R_{-}$ for HNC/MSA$_{RPM}$, and 
$d_{-\,+}=d_{+\,-}=0$ for URMGC and MGC.}
\label{fig_z0z1_c2} 
\end{minipage} 
\end{center} 
\end{figure}

For the theories mentioned in the foregoing the dependence of $\psi _{0}$
and $\psi _{IHP}$ on the surface charge $\sigma _{0}$ is given in Figs. 
\ref{fig_z0z1_c2}(a) and \ref{fig_z0z1_c2}(b), respectively, for a macroion 
with $D=160$ \AA\ and $\sigma_{0}\geq 0$ and a 1:1, 1 M electrolyte with 
$R_{-}=4.25$~\AA\ and $R_{+}=8.5$~\AA. In those figures, 
for any non-zero value of $\sigma _{0}$, visible
quantitative discrepancies are seen when the URMGC and HNC/MSA$_{PM}$
potential curves are contrasted, or when the same is done with the MGC and
HNC/MSA$_{RPM}$ ones. Remarkably, the maximum coincidence between the pair
of URMGC and HNC/MSA$_{PM}$ curves of $\psi _{0}(\sigma _{0})$ (and of 
$\psi _{IHP}(\sigma _{0})$) occurs precisely at the point of zero charge. In
a similar way, the MGC and HNC/MSA$_{RPM}$ values of $\psi _{0}$\ (and of 
$\psi _{IHP}$) converge at $\sigma _{0}=0$, as it is known from the studies of 
the charge-symmetric RPM-EDL \cite{Ca02,Lo01,Go01,Go02}
(for the charge-asymmetric case see \cite{Lo09,Bo03,Va02,Bh03,To05}). On the
other hand, for a given value of $\sigma _{0}$, the differences between the
results of a Poisson-Boltzmannian theory and the corresponding 
non-punctual HNC/MSA 
equation for $\psi _{0}$ and $\psi _{IHP}$ grow when size asymmetry is taken
into account, e.g. at $\sigma _{0}=0.1$~C/m$^{2}$, 
$(\psi _{0})^{MGC}-(\psi_{0})^{HNC/MSA_{RPM}}=10.04~$mV whereas 
$(\psi _{0})^{URMGC}-(\psi_{0})^{HNC/MSA_{PM}}=24.11~$mV 
(see Fig. \ref{fig_z0z1_c2}(a)), 
and $(\psi _{IHP})^{MGC}-(\psi_{IHP})^{HNC/MSA_{RPM}}=10.04~$mV whereas 
$(\psi _{IHP})^{URMGC}-(\psi_{IHP})^{HNC/MSA_{PM}}=24.12~$mV 
(see Fig. \ref{fig_z0z1_c2}(b)). Also, from 
Fig. \ref{fig_z0z1_c2} it is
confirmed that the potential curves of URMGC approach asymptotically to
those of MGC when $\sigma _{0}$ increases, as first pointed out by Valleau
and Torrie \cite{Va01}. Contrastingly, the $\psi _{0}(\sigma _{0})$ and 
$\psi_{IHP}(\sigma _{0})$ curves obtained from HNC/MSA$_{PM}$ and 
HNC/MSA$_{RPM}$
exhibit a clear separation for all surface charges, which means that 
\textit{counterions do not dominate in the SEDL} or, else, that the size of 
the coions matters
even for large $\sigma _{0}$. From these observations it is infered that:
i) an essentially punctual theory (MGC or URMGC) is not valid to describe
the SEDL under high coupling conditions in which the size of the ions is 
relevant for all
their interactions (ion-ion or colloid-ion), except maybe in the
neighborhood of $\sigma _{0}=0$, and, two important points, ii) that including
size-asymmetry in the model of an EDL is determinant since it exacerbates
the steric interionic effects previously found in the RPM-EDL and iii) that
the properties of the EDL are not totally determined by the counterions.

\begin{figure}[htbp] 
\begin{center} 
\begin{minipage}{\linewidth} 
{\includegraphics[angle=0.0, width=\linewidth]{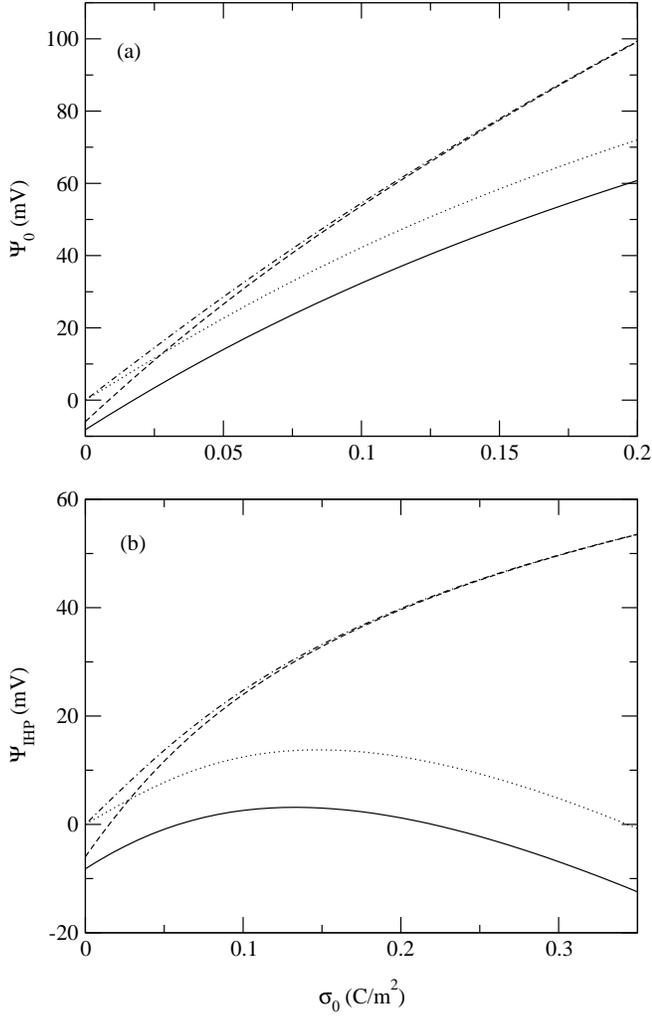}}
\caption{
The same as in Fig. \ref{fig_z0z1_c2} but for a 2:2, 0.5 M electrolyte.} 
\label{fig_z0z1_c6} 
\end{minipage} 
\end{center} 
\end{figure}

Expectedly, it is found that all the referred phenomena detected in
monovalent electrolytes are more pronounced for divalent ions, as it is
evidenced in Figs. \ref{fig_z0z1_c6}(a) and \ref{fig_z0z1_c6}(b) where 
the associated HNC/MSA$_{PM}$, HNC/MSA
$_{RPM}$, URMGC and MGC results for $\psi _{0}(\sigma _{0})$ and $\psi
_{IHP}(\sigma _{0})$ are reported\ for a 2:2, 0.5 M electrolyte with 
$R_{-}=4.25$~\AA\ and $R_{+}=8.5$~\AA, around a
macrosphere of diameter $D=160$ \AA\ and positive charge. For instance, now
the discrepancies between the MEPs of PB and HNC/MSA have increased in such
a way that even the potentials of zero charge, 
$\psi _{0}(\sigma_{0}=0)$ and $\psi _{IHP}(\sigma _{0}=0)$, for URMGC and 
HNC/MSA$_{PM}$ do not coincide. Once more,
URMGC has MGC as a limit for $\sigma _{0}\rightarrow \infty $ whereas
HNC/MSA$_{PM}$ and HNC/MSA$_{RPM}$ do not merge, notwithstanding, this time
the following very interesting feature shows up: the $\psi _{IHP}(\sigma _{0})$
relationship for HNC/MSA$_{RPM}$ exhibits the usual behavior documented in
prior studies of positively charged colloids in 2:2 solutions 
\cite{Go02,Go03}, i.e. it
begins positive, reaches a maximum and becomes negative only at very high
surface charges ($\sigma _{0}\simeq 0.344~$C/m$^{2}$), whilst, in
contrast, $\psi _{IHP}(\sigma _{0})$ for HNC/MSA$_{PM}$ behaves differently,
that is, \textit{starts negative at low }$\sigma _{0}$, then regains
``normal'' positive values, experiences a maximum and, finally, becomes
negative again at $\sigma _{0}\simeq 0.22~$C/m$^{2}$, much before 
HNC/MSA$_{RPM}$ does. This new fact could be consequential in electrokinetics 
since
the signs of the MEP at the IHP and of the electrophoretic mobility of a
colloid, $\mu $, frequently coincide (due to the identification $\zeta
=\psi _{IHP}$), from which \textit{it should be possible to observe
experimentally a macroion in a multivalent medium having a reentrant 
mobility, i.e. with inversion
of }$\mu$ \textit{at low }$\sigma _{0}$\textit{, a normal sign for
intermediate charges, and a posterior return to a inverted }$\mu $ 
\textit{for high }$\sigma _{0}$.

A more detailed examination of the size asymmetry consequences, already
discussed at the level of the potential-charge relationship, can be made in
terms of the SEDL structure. Thence, in the next 
figures, Figs. \ref{new_gs_c2} and \ref{new_gs_c2_3}, we
proceed to analyze the behavior of the ionic profiles of univalent solutions
described by the MGC, URMGC, HNC/MSA$_{RPM}$ and HNC/MSA$_{PM}$ formalisms.
To establish a connection with our previous MEP curves, the RDFs were
obtained for a 1:1, 1 M electrolyte, with $R_{-}=4.25$~\AA\ and 
$R_{+}=8.5$~\AA, dissolving a macroion of size $D=160$~\AA\ and surface 
charges $\sigma _{0}=0.08356~$C/m$^{2}$ (in Fig. \ref{new_gs_c2}) and 
$\sigma _{0}=0.3004~$C/m$^{2}$
(in Fig. \ref{new_gs_c2_3}). For each colloidal charge, the HNC/MSA$_{RPM}$ 
and HNC/MSA$_{PM}$ data are compared in the main panels of the 
figures, meanwhile, in
the insets the MGC and URMGC plots are collated, to evaluate separately
the effect of varying $\sigma _{0}$ on the distributions $g_{i}(r)$ when
size asymmetry is incorporated into HNC/MSA or PB theories. From the
insets it is manifest the almost complete convergence between the 
\textit{monotonic} URMGC and MGC RDFs for growing $\sigma _{0}$, as it could be
awaited from the preceding results for $\psi _{0}(\sigma _{0})$ and 
$\psi_{IHP}(\sigma _{0})$ \cite{Va01}. In any case, some distinctions
between the URMGC\ and MGC profiles must naturally persist for all $\sigma
_{0}$ (with the only exception of the unphysical value $\sigma _{0}=\infty$), 
due to the severe condition $g_{2}(r)=0$ for $r<(D+R_{2})/2$. Such
restriction demands a finite discontinuity in $g_{2}(\frac{D+R_{2}}{2})$ for
URMGC and thus, for $0\leq \sigma _{0}<\infty $, impedes a perfect
coincidence between the RDFs of URMGC and MGC in the vicinity of the
colloid surface. In other words, the total equivalence of URMGC and MGC is
possible just in the limit of infinite surface charge, since only then the
infinite and predominant electrostatic repulsion from the macroparticle will
force the coions in the MGC description to be completely absent from the
region $(D+R_{1})/2\leq r<(D+R_{2})/2$, even if their punctual nature would
allow them to be there. To be more explicit, if we define the 
abbreviations $C_{-}=(D+R_{-})/2$ and $C_{+}=(D+R_{+})/2$, an inspection of the
1:1 counterion contact values, $g_{-}(C_{-})$, in Table \ref{t2} confirms the
existence of small but non-zero differences between the URMGC and MGC RDFs
close to the macroparticle. Moreover, an important piece of information
arises from Table \ref{t2} and the insets of
Figs. \ref{new_gs_c2} and \ref{new_gs_c2_3} for systems with 
$\sigma _{0}\geq 0$ and $R_{-}<R_{+} $ when $\sigma _{0}$ is enlarged, namely
that the URMGC counterion contact RDF tends uniformly to that of MGC from
below ($g_{-}^{URMGC}(C_{-})\uparrow g_{-}^{MGC}(C_{-})$), and that the
URMGC coion RDF at $C_{+}$ goes uniformly to that of MGC from above 
($g_{+}^{URMGC}(C_{+})\downarrow g_{+}^{MGC}(C_{+})$). In a more global way,
it is noticed therein that, $\forall $ $r$, $g_{-}^{URMGC}(r)\uparrow
g_{-}^{MGC}(r)$ and $g_{+}^{URMGC}(r)\downarrow g_{+}^{MGC}(r)$ as the
colloidal charge grows, or, equivalently, that \textit{the URMGC and MGC
ionic profiles converge when }$\sigma _{0}\rightarrow \infty$ \textit{but
they never cross mutually}.

On the other hand, the comportment of the RDFs for HNC/MSA$_{PM}$ and
HNC/MSA$_{RPM}$ in the main panels of Figs. \ref{new_gs_c2} and 
\ref{new_gs_c2_3} contrasts with the PB
picture at the insets. For our 1:1 electrolyte with counterions smaller than
coions, the ionic distributions of HNC/MSA\ in the PM and RPM already
exhibit a nonmonotonic behavior and charge inversion (in the case of RPM
these features are unapparent due to the scale). Besides, from those graphs
we arrive to one of the main conclusions of this work, that \textit{size
asymmetry enhances significantly the nonmonotonical characteristics (e.g.
oscillations and charge inversion) of the SEDL}. In relation with the
conduct of the SEDL structure for high $\sigma _{0}$, 
\textit{the HNC/MSA}$_{PM}$\textit{ionic densities do not have 
the HNC/MSA}$_{RPM}$ \textit{ones as a limit} and, in fact, from Table 
\ref{t2} it is seen that the contact
values of the counterion RDF in HNC/MSA$_{PM}$ exceed appreciably those of
HNC/MSA$_{RPM}$ even for large values of $\sigma _{0}$. Additionally, 
\textit{for a constant surface charge density, 
the HNC/MSA}$_{PM}$\textit{RDFs for counterions and coions fluctuate more 
strongly and have a steeper 
slope than the corresponding to HNC/MSA}$_{RPM}$\textit{, and,
consequently, crossings between the profiles for PM and RPM do occur}.

\begin{figure}[htbp] 
\begin{center} 
\begin{minipage}{\linewidth} 
{\includegraphics[angle=0.0, width=\linewidth]{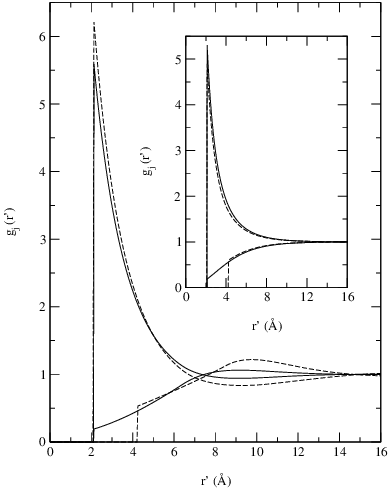}}
\caption{SEDL\ radial distribution functions for a 2:2, 0.5 M electrolyte
with $R_{-}=4.25$~\AA\ and $R_{+}=8.5$~\AA\ next to a colloid of diameter 
$D=160$ \AA\ and surface charge density $\sigma _{0}=0.08356$ C/m$^{2}$. In
the main panel the solid and dashed lines stand for HNC/MSA$_{RPM}$ and
HNC/MSA$_{PM}$, respectively, whereas in the inset the solid and dashed
lines correspond to MGC and URMGC, respectively. The colloid-ion and ion-ion
closest approach distances used in each theory are the same as in 
Fig. \ref{fig_z0z1_c2}. 
}
\label{new_gs_c6} 
\end{minipage} 
\end{center} 
\end{figure}

\begin{figure}[htbp] 
\begin{center} 
\begin{minipage}{\linewidth} 
{\includegraphics[angle=0.0, width=\linewidth]{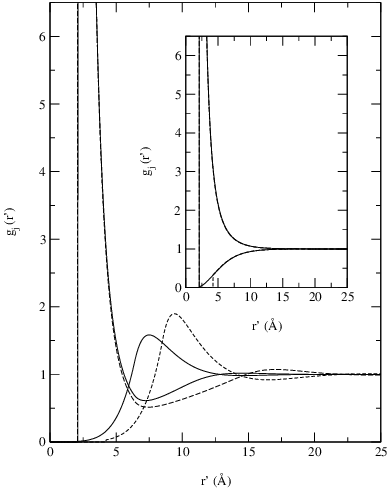}}
\caption{
The same as in Fig. \ref{new_gs_c6} but for a surface charge density 
$\sigma_{0}=0.3004$ C/m$^{2}$.
} 
\label{new_gs_c6_3} 
\end{minipage} 
\end{center} 
\end{figure}

For divalent suspensions the RDFs for a 2:2, 0.5 M electrolyte, with 
$R_{-}=4.25$~\AA\ and $R_{+}=8.5$~\AA, and a macrosphere of $D=160$ \AA\ are
displayed in Fig. \ref{new_gs_c6} for a surface charge 
$\sigma _{0}=0.08356~$C/m$^{2}$ and
in Fig. \ref {new_gs_c6_3} for $\sigma _{0}=0.3004~$C/m$^{2}$. Again, in the
main panels
HNC/MSA$_{PM}$ and HNC/MSA$_{RPM}$ are plotted and in the insets URMGC and
MGC. In general, the situation described for the RDFs of monovalent
electrolytes is repeated here: URMGC\ and MGC are monotonic, whereas 
HNC/MSA$_{PM}$ and HNC/MSA$_{RPM}$ oscillate, and, for large $\sigma _{0}$, 
URMGC
converge to MGC with no intersections between them and HNC/MSA$_{RPM}$ is
not the limit of HNC/MSA$_{PM}$. Furthermore, Table \ref{t2} ratifies that the
counterion contact values, $g_{-}(C_{-})$ are larger for HNC/MSA$_{PM}$ than
for HNC/MSA$_{RPM}$, and Figs. \ref{new_gs_c6} and \ref {new_gs_c6_3} 
evince that the ionic profiles of the
former theory wave more intensely than the ones of the latter. Rewardingly,
this last pair of structural features of HNC/MSA accords with the
ARHNC information for a 2:2 planar EDL obtained by Greberg and
Kjellander \cite{Gr02}.

As a clear example of how the microscopic structure determines other
properties of the EDL, on the grounds of the discussion above we will
formulate a rationale for the contrasting high-$\sigma _{0}$ behavior of the 
$\psi _{0}(\sigma _{0})$ and $\psi _{IHP}(\sigma _{0})$ curves obtained from
PB and HNC/MSA. Aiming for that, let us first condense some of our recent
findings in a more convenient form. For size-asymmetric SEDL systems with
counterions smaller than coions, if $g_{i}(r;\sigma _{0})$ represents an
ionic profile calculated at a given colloidal charge, (i) when $\sigma _{0}$
is varied the URMGC RDFs approach from one side to those of MGC, and the
separation between the URMGC and MGC profiles decreases progressively as 
$\sigma _{0}$ is enlarged, i.e. \textit{there are no crossings between the }
$g_{i}^{URMGC}(r;\sigma _{0})$\textit{ and }$g_{i}^{MGC}(r;\sigma _{0})$
\textit{functions, and the MGC RDFs bound those of URMGC when }
$\sigma_{0}\rightarrow \infty $, and (ii) the RDFs at contact for counterions 
in HNC/MSA$_{PM}$ have larger values than in HNC/MSA$_{RPM}$, and the ionic
densities of HNC/MSA$_{PM}$ have a more pronounced slope than the ones of 
HNC/MSA$_{RPM}$
do, i.e. $g_{i}^{HNC/MSA_{PM}}(r;\sigma _{0})$\textit{ and }
$g_{i}^{HNC/MSA_{RPM}}(r;\sigma _{0})$\textit{\ present intersections, and,
for large }$\sigma _{0}$\textit{, the RDFs of HNC/MSA}$_{PM}$
\textit{ are not bounded nor have HNC/MSA}$_{RPM}$\textit{\ as a limit}. It 
must be noted
that in all the previous statements the comparisons between the theories are
performed \textit{at fixed }$\sigma _{0}$. If we restate explicitely the 
$\sigma _{0}$-integral, Eq. (\ref{q_0}), for a binary electrolyte:

\begin{equation}
\sigma _{0}=\left( \frac{e\,\rho_{+}\,z_{+}}{\left( D/2\right) ^{2}}\right) \int_{D/2}^{\infty }\left[ g_{-}(t)-g_{+}(t)\right] \,t^{2}\,dt,
\end{equation}

\noindent then, for $\sigma _{0}\geq 0$, this surface charge integral is
proportional to the difference between the areas under the $r^{2}$-functions 
$f_{-}(r;\sigma _{0})\equiv g_{-}(r;\sigma _{0})\,r^{2}$ and $f_{+}(r;\sigma
_{0})\equiv g_{+}(r;\sigma _{0})\,r^{2}$, for counterions and coions,
respectively. In such terms, if we consider the data in each of the 
Figs. \ref{new_gs_c2} to \ref{new_gs_c6_3}, irrespective of the theory, 
every of the four pairs of
counterion/coion $r^{2}$-functions, $g_{-}(r;\sigma _{0})\,r^{2}$ and 
$g_{+}(r;\sigma _{0})\,r^{2}$, comprises a constant area. This rephrasing of
the total electroneutrality condition as a geometrical constraint for the
RDFs can now be employed to clarify why the URMGC and MGC potential-charge
curves converge for $\sigma _{0}\rightarrow \infty $, whereas the
corresponding to HNC/MSA$_{PM}$ and HNC/MSA$_{RPM}$ do not, or else, why the
counterions do not dominate in a description of the size-asymmetric SEDL
that surpasses the classical PB equation (for instance, HNC/MSA).

\begin{figure}[htbp] 
\begin{center} 
\begin{minipage}{\linewidth} 
{\includegraphics[angle=0.0, width=\linewidth]{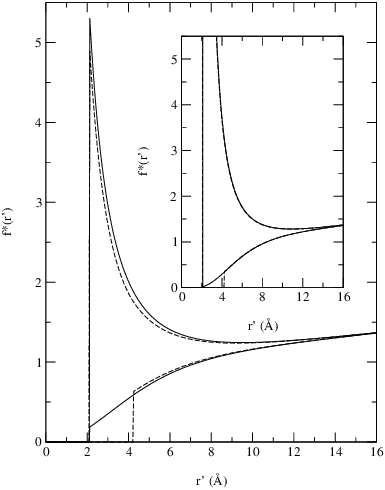}}
\caption{SEDL reduced 
profiles $f^\ast(r^{\prime})=f((D/2)+r^{\prime})/C_{-}^{2}$ for 
a 2:2, 0.5 M with $R_{-}=4.25$~\AA\ and $R_{+}=8.5$~\AA\ 
around a colloid of diameter $D=160$ \AA\ . The solid and dashed lines
are for MGC and URMGC, respectively. In the main panel the surface charge
density is $\sigma _{0}=0.08356$ C/m$^{2}$, and in the inset is $\sigma
_{0}=0.3004$ C/m$^{2}$. The colloid-ion and ion-ion closest approach
distances used in each theory are the same as in Fig. \ref{fig_z0z1_c2}. 
} 
\label{pb_gr2} 
\end{minipage} 
\end{center} 
\end{figure}

\begin{figure}[htbp] 
\begin{center} 
\begin{minipage}{\linewidth} 
{\includegraphics[angle=0.0, width=\linewidth]{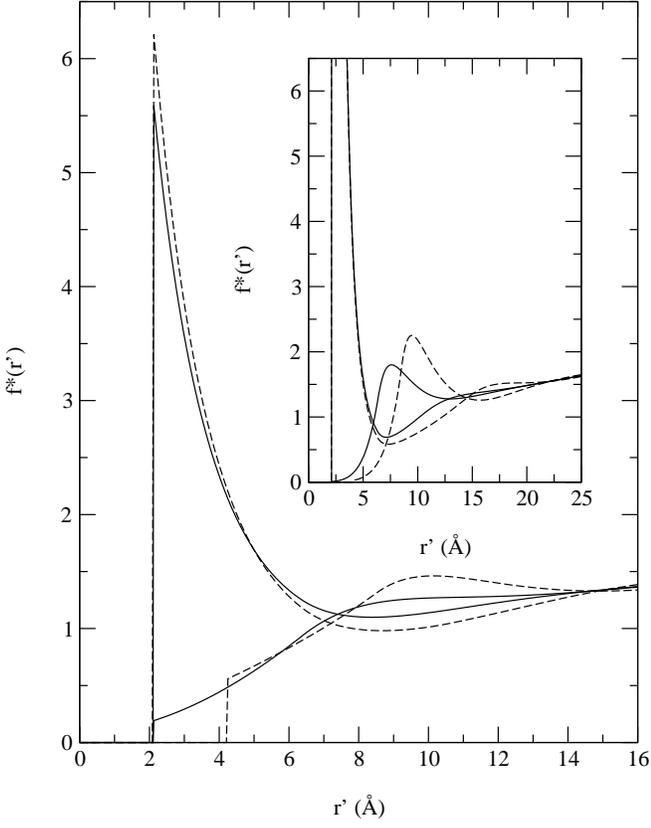}}
\caption{SEDL reduced 
profiles $f^\ast(r^{\prime})=f((D/2)+r^{\prime})/C_{-}^{2}$
for a 2:2, 0.5 M with $R_{-}=4.25$~\AA\ and $R_{+}=8.5$ \AA\ around a colloid 
of diameter $D=160$ \AA. The solid and dashed lines
are for HNC/MSA$_{RPM}$ and HNC/MSA$_{PM}$, respectively. In the main panel
the surface charge density is $\sigma _{0}=0.08356$ C/m$^{2}$, and in the
inset is $\sigma _{0}=0.3004$ C/m$^{2}$. The colloid-ion and ion-ion closest
approach distances used in each theory are the same as in 
Fig. \ref{fig_z0z1_c2}. 
} 								
\label{hm_gr2} 
\end{minipage} 
\end{center} 
\end{figure}

To examine the fulfillment of the constant area restriction as $\sigma _{0}$
is enlarged, in Figs. \ref{pb_gr2} and \ref{hm_gr2} the $r^{2}$-\ functions,
 $f_{-}(r;\sigma_{0})\equiv g_{-}(r;\sigma _{0})\,r^{2}$ and 
$f_{+}(r;\sigma _{0})\equiv g_{+}(r;\sigma _{0})\,r^{2}$, 
resulting from the four theories in
consideration: URMGC, MGC, HNC/MSA$_{PM}$ and HNC/MSA$_{RPM}$, are plotted
(in reduced form)
for a 2:2, 0.5 M electrolyte of ionic sizes $R_{-}=4.25$~\AA\ and 
$R_{+}=8.5$~\AA , and a macroion of diameter $D=160$ \AA\ bearing two 
different charge
densities, $\sigma _{0}=0.08356~$C/m$^{2}$ (in the main sections of the
figures) and $\sigma _{0}=0.3004~$C/m$^{2}$ (in the insets). For the URMGC
and MGC data of the lower charge, presented in the main panel of 
Fig. \ref{pb_gr2}, 
a comparison between the $r^{2}$-functions for coions immediately shows that
some area below the URMGC coion function is ``lost'' in the region $C_{-}\leq
r<C_{+}$, owing to the hard-core condition. On the other hard, since the
RDFs of URMGC and MGC do not intersect, and remembering the PB
reciprocity property for $n:n$ electrolytes in the region $r\geq C_{+}$:

\begin{equation}
g_{+}^{PB}(r)\,g_{-}^{PB}(r)=1,
\end{equation}

\noindent it is straightforwardly concluded that the ionic profiles 
$g_{-}^{URMGC}(r;\sigma _{0})$ and $g_{+}^{URMGC}(r;\sigma _{0})$ are
necessarily located in the region enclosed by $g_{-}^{MGC}(r;\sigma _{0})$
and $g_{+}^{MGC}(r;\sigma _{0})$ (i.e. $\forall $ $r$, $g_{-}^{URMGC}(r;%
\sigma _{0})<g_{-}^{MGC}(r;\sigma _{0})$ and $g_{+}^{URMGC}(r;\sigma
_{0})>g_{+}^{MGC}(r;\sigma _{0})$) as the unique way in which the coion $%
r^{2}$-\ function of URMGC can recover area in $r\geq C_{+}$ to assure that
the difference functions, $g_{d}(r;\sigma _{0})\,r^{2} = \left[ g_{-}(r;\sigma _{0})-g_{+}(r;\sigma _{0})\right] \,r^{2}$, of URMGC and MGC 
preserve a constant area. Additionally, when $\sigma _{0}$ grows 
(see inset of Fig. \ref{pb_gr2}) the MGC
coion RDF at $r=C_{+}$ goes down and the coion $r^{2}$-function of URMGC
has a minor deficit of area in $C_{-}\leq r<C_{+}$, from which the separation
between the URMGC and MGC RDFs diminishes and, eventually, disappears 
at $\sigma _{0}=\infty $. By
contrast, in the HNC/MSA case (Fig. \ref{hm_gr2})
there is also a lost of area in the zone 
$C_{-}\leq r<C_{+}$ for the coion $r^{2}$-function of HNC/MSA$_{PM}$,
however the possibilty of crossings between the ionic profiles of 
HNC/MSA$_{PM}$ and HNC/MSA$_{RPM}$, stemming from the enhancement of the
non-monotonicity of the RDFs caused by the size-asymmetry, does not
obligate the coincidence of $g_{i}^{HNC/MSA_{PM}}(r;\sigma _{0})$\textit{\ }
and\textit{\ }$g_{i}^{HNC/MSA_{RPM}}(r;\sigma _{0})$ for large $\sigma _{0}$, 
even if the difference
functions of both theories have the same area. It is from these reasonings
that the convergence, for $\sigma _{0}\rightarrow \infty $, between the
structural properties of URMGC and MGC and the non-convergence for those of
HNC/MSA$_{PM}$ and HNC/MSA$_{RPM}$ can be understood. A similar analysis
done for 1:1 systems (not shown) confirms all the arguments exposed.

Summing up, when the value of $\sigma _{0}$ rises, the difference function 
$g_{d}(r;\sigma _{0})\,r^{2}=\left[ g_{-}(r;\sigma _{0})-g_{+}(r;\sigma _{0})\right] \,r^{2}$\ for each of the MGC, URMGC, HNC/MSA$_{RPM}$ and 
HNC/MSA$_{PM}$ approximations can accomplish the constraint of a constant area
in two ways. On the one hand, when $\sigma _{0}\rightarrow \infty $, the 
$r^{2}$-integrals (or second moments) of $g_{d}^{URMGC}(r;\sigma _{0})$ and 
$g_{d}^{MGC}(r;\sigma _{0})$ remain exactly equal whereas all the rest of
integrals of such functions (including the $\psi _{0}$- and 
$\psi _{IHP}$-integrals; cf. Eqs. ((\ref{pot_0}) and (\ref{pot_ihp})) become 
progressively alike, since the
RDFs of URMGC and MGC approach asymptotically. On the other, when 
$\sigma_{0}$ increases, the pair $g_{d}^{HNC/MSA_{PM}}(r;\sigma _{0})$ and 
$g_{d}^{HNC/MSA_{RPM}}(r;\sigma _{0})$ have the same $r^{2}$-integral but
other integrals (such as the $\psi _{0}$- and $\psi _{IHP}$-integrals)
become different, since the HNC/MSA$_{PM}$ and HNC/MSA$_{RPM}$ ionic
profiles do not merge. This cogently explains why the 
$\psi _{0}(\sigma_{0}) $ and $\psi _{IHP}(\sigma _{0})$ curves for URMGC and 
MGC converge for large $\sigma _{0}$ and those for HNC/MSA$_{PM}$ and 
HNC/MSA$_{RPM}$ remain separated.

\begin{figure}[htbp] 
\begin{center} 
\begin{minipage}{\linewidth} 
{\includegraphics[angle=0.0, width=\linewidth]{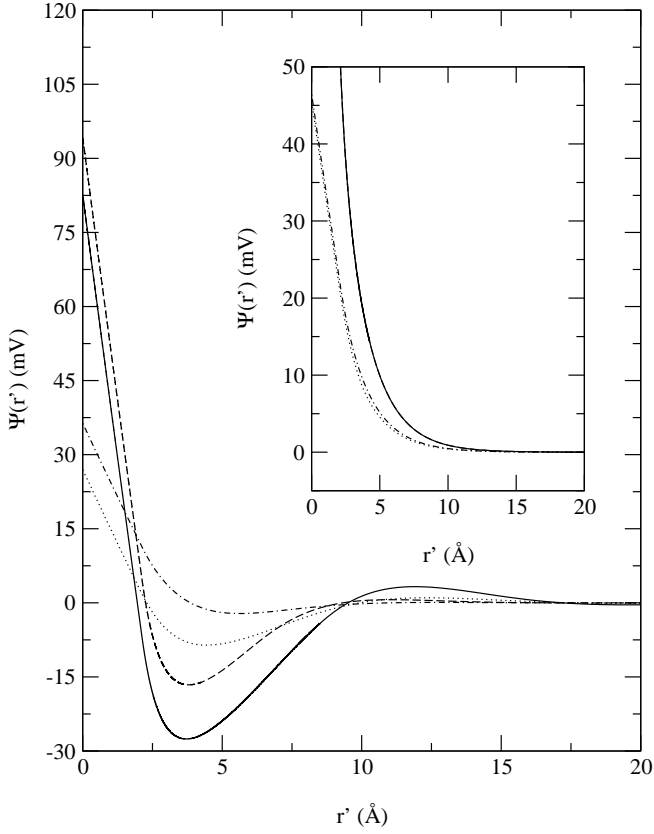}}
\caption{SEDL mean electrostatic potential as a function of the distance to
the macroparticle's surface for a 2:2, 0.5 M with $R_{-}=4.25$~\AA\ and 
$R_{+}=8.5$~\AA\ around a colloid of diameter $D=160$ \AA\ . In the main panel
the dotted line is for HNC/MSA$_{PM}$ and $\sigma _{0}=0.08356$ C/m$^{2}$,
the solid line is for HNC/MSA$_{PM}$ and $\sigma _{0}=0.3004$ C/m$^{2}$, the
dot-dashed is for HNC/MSA$_{RPM}$ and $\sigma _{0}=0.08356$ C/m$^{2}$, and
the dashed line is for HNC/MSA$_{RPM}$ and $\sigma _{0}=0.3004$ C/m$^{2}$.
In the inset the dotted line is for URMGC and $\sigma _{0}=0.08356$ C/m$^{2}$,
the solid line is for URMGC and $\sigma _{0}=0.3004$ C/m$^{2}$, the
dot-dashed is for MGC and $\sigma _{0}=0.08356$ C/m$^{2}$, and the dashed
line is for MGC and $\sigma _{0}=0.3004$ C/m$^{2}$. The colloid-ion and
ion-ion closest approach distances used in each theory are the same as in 
Fig. \ref{fig_z0z1_c2}
} 
\label{new_vs_c6} 
\end{minipage} 
\end{center} 
\end{figure}

\begin{figure}[htbp] 
\begin{center} 
\begin{minipage}{\linewidth} 
{\includegraphics[angle=0.0, width=\linewidth]{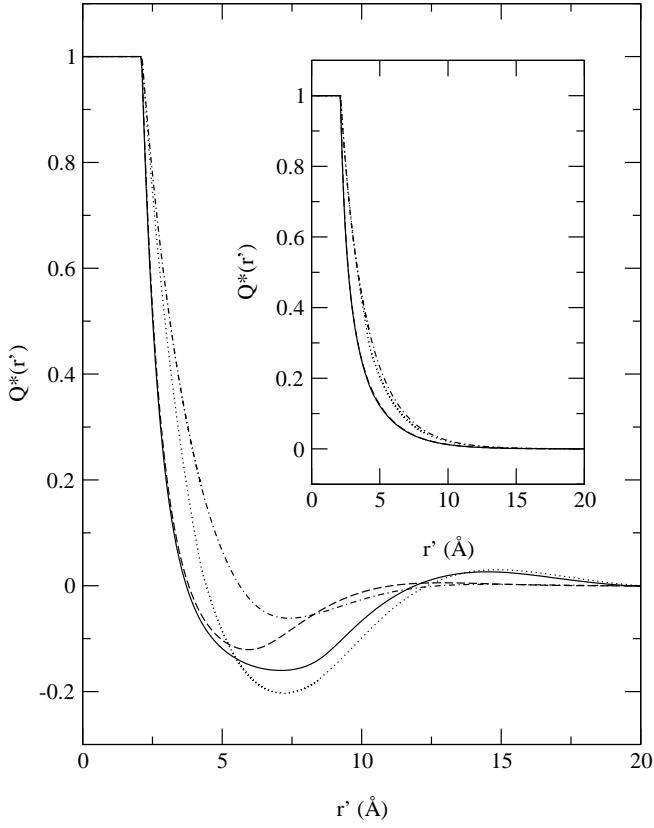}}
\caption{SEDL cumulative reduced charge as a function of the distance to the
macroparticle's surface for a 2:2, 0.5 M with $R_{-}=4.25$~\AA\ and 
$R_{+}=8.5$~\AA\ around a colloid of diameter $D=160$ \AA\ . In the main panel
the dotted line is for HNC/MSA$_{PM}$ and $\sigma _{0}=0.08356$ C/m$^{2}$,
the solid line is for HNC/MSA$_{PM}$ and $\sigma _{0}=0.3004$ C/m$^{2}$, the
dot-dashed is for HNC/MSA$_{RPM}$ and $\sigma _{0}=0.08356$ C/m$^{2}$, and
the dashed line is for HNC/MSA$_{RPM}$ and $\sigma _{0}=0.3004$ C/m$^{2}$.
In the inset the dotted line is for URMGC and $\sigma _{0}=0.08356$ C/m$^{2}$,
the solid line is for URMGC and $\sigma _{0}=0.3004$ C/m$^{2}$, the
dot-dashed is for MGC and $\sigma _{0}=0.08356$ C/m$^{2}$, and the dashed
line is for MGC and $\sigma _{0}=0.3004$ C/m$^{2}$. The colloid-ion and
ion-ion closest approach distances used in each theory are the same as in 
Fig. \ref{fig_z0z1_c2}.
} 
\label{new_qs_c6} 
\end{minipage} 
\end{center} 
\end{figure}

To complete this study of the PM-SEDL we switch our attention to the 
MEP and CRC profiles, $\psi (r)$ and $Q^{\ast }(r)$, respectively, to gain 
more insight into the
comportment of the diffuse EDL in terms of its neutralization (or charge
screening) capacity. In Figs. \ref{new_vs_c6} and \ref{new_qs_c6} 
the MEP and CRC functions
corresponding to the 2:2 data in Figs. \ref{new_gs_c6} and \ref{new_gs_c6_3}
are portrayed, respectively.
We have preferred to graph the divalent and not the univalent results since
the effects to be discussed have qualitative similarities in the 2:2 and 1:1
systems but are more marked in the former case. In respect to the $\psi (r)$
profiles, the URMGC and MGC MEPs included in the inset of Fig. \ref{new_vs_c6}
confirm the monotonic character of the PB solutions and the convergence between
URMGC and MGC for $\sigma _{0}\rightarrow \infty $. In turn, all the
HNC/MSA MEP profiles in the main panel of Fig. \ref{new_vs_c6} oscillate and 
present ample regions of inverted potentials or even show an alternation of 
signs (see, for example, HNC/MSA$_{PM}$ for $\sigma _{0}=0.3004~$C/m$^{2}$).
However, the most salient feature in both the PB and HNC/MSA results of $%
\psi (r)$ is that, in the proximities of the colloid, an SEDL description
in which size asymmetry is assumed exhibits lower MEP values compared to the
corresponding potentials of the size-symmetric version of the same theory,
i.e. $\psi ^{HNC/MSA_{PM}}(r;\sigma _{0})<\psi ^{HNC/MSA_{RPM}}(r;\sigma
_{0})$ and $\psi ^{URMGC}(r;\sigma _{0})<\psi ^{MGC}(r;\sigma _{0})$, for $%
(r-D/2)\lesssim 7.5\,$\AA . A complementary and more quantitative
estimation of this can be obtained from the values collected in Table \ref{t3}.
The observed behavior of the $\psi (r)$ functions close to the 
surface implies that a
size-asymmetric diffuse EDL (either of HNC/MSA$_{PM}$ or URMGC) neutralizes
more efficiently the native macroion charge than that of a size-symmetric
system (of HNC/MSA$_{RPM}$ or MGC). This generalized dominance of the
neutralization power of an unequally-sized electrolyte is remarkable since
in the PB approaches there are less counterions close to the macroparticle
in URMGC than in MGC, in other words, the counterion contact RDFs comply
with $g_{-}^{URMGC}(C_{-};\sigma _{0})<g_{-}^{MGC}(C_{-};\sigma _{0})$,
whilst in the HNC/MSA theories the contact number of counterions in 
HNC/MSA$_{PM}$ exceeds the prediction of HNC/MSA$_{RPM}$, i.e. 
$g_{-}^{HNC/MSA_{PM}}(C_{-};\sigma _{0})>g_{-}^{HNC/MSA_{RPM}}(C_{-};\sigma
_{0})$. Such ``anomaly'' can be justified nicely by the obligated absence of
coions in the zone $C_{-}\leq r<C_{+}$ for the HNC/MSA$_{PM}$ and URMGC
theories, which, irrespective of the PB or HNC/MSA approach, allows the
counterions to compensate the colloidal charge in a better way than it is
done in HNC/MSA$_{RPM}$ or MGC, theories where the coions can be indeed
present in $C_{-}\leq r<C_{+}$ due to the condition $d_{+}=d_{-}$. This, 
however, does not explain the additional neutralization power of
HNC/MSA$_{PM}$ over URMGC. For HNC/MSA$_{PM}$ the reason is the increase
of the excluded volume, which in turns promotes the coions and counterions
adsorption, due to entropic effects, as it has been discussed by 
Jim\'enez-\'Angeles and Lozada-Cassou \cite{Ji01}. These last effects also 
explicate the increase of the counterion  adsorption of the HNC/MSA$_{PM}$
over HNC/MSA$_{RPM}$ and, consequently, the enhancement of the charge 
screening in the PM-SEDL.

The behavior of the neutralization capacity detected via the mean
electrostatic potential profiles fits very well with the information
provided by the cumulative reduced charges. For instance, in the inset of
Fig. \ref{new_qs_c6} it can be verified that the $Q^{\ast }(r)$ functions 
for URMGC and
MGC are monotonic and do not display any change of sign, contrasting with
the main panel of the same figure where HNC/MSA$_{PM}$ and HNC/MSA$_{RPM}$
have CRC curves with oscillations and appreciable charge reversion.
Additionally, a genuine and clearer manifestation of the enhanced
neutralization capacity of size-asymmetric models over the symmetrical ones
can be gathered directly from the full set of CRC curves in 
Fig. \ref{new_qs_c6}. There,
it is visible the faster decay in the cumulative charges corresponding to
HNC/MSA$_{PM}$ and URMGC with respect to those of HNC/MSA$_{RPM}$ and MGC,
i.e. $Q^{\ast \,HNC/MSA_{PM}}(r;\sigma _{0})<Q^{\ast
\,HNC/MSA_{RPM}}(r;\sigma _{0})$ and $Q^{\ast \,URMGC}(r;\sigma
_{0})<Q^{\ast \,MGC}(r;\sigma _{0})$, for $(r-D/2)\lesssim 10\,$\AA.

Notably, for a given $\sigma _{0}$, among all the reviewed theories, the
HNC/MSA$_{PM}$ formalism is always associated to the strongest fluctuations
and to the more accentuated sign inversions in $\psi (r)$ and to the
largest neutralization and charge reversal effects in $Q^{\ast }(r)$. This
fact, together with the rest of findings revealed by the analysis of the 
$\psi (r)$ and $Q^{\ast }(r)$ profiles, conforms an extra and compelling
testimony to the relevance of using size-asymmetric models when a more
faithful description of colloidal suspensions is sought.

To end, we want to point out that, for our $n:n$ systems, at the distances
where the RDFs show a charge inversion the CRCs go through a minimum or
maximum, as it was noted previously by Wang \textit{et al}. in the RPM-EDL
for cylindrical geometry \cite{Wa01}. This signifies that, for a
valence-symmetric RPM or PM electrolyte, charge inversion implies charge
reversal. What is more, from basic electrostatics for the SEDL, it is 
possible to
prove that $\frac{d\psi (r)}{dr}=-\frac{Q(r)}{\epsilon \,r^{2}}$ (similar
equations for other geometries are given in Refs. \cite{Wa01,Ji01}), thus, the
distances at which the CRC is zero correspond precisely to those where the
mean electrostatic potential presents extremum values, whence, the existence
of oscillations in $Q^{\ast }(r)$ should not be enough to produce
fluctuations in $\psi (r)$, i.e. a change of sign in the CRC is indeed
required.

\section{CONCLUSIONS} 

In this work a survey of the size-asymmetric spherical electrical
double layer in the primitive model was performed by using the HNC/MSA
integral equation. After the correctness of the HNC/MSA description was
attested by comparing our numerical results with simulations and previous
ARHNC data (in the planar limit), we have carried out a study of
the structural and charge-potential relationship for archetypal cases of 1:1
and 2:2 systems with a size-asymmetry ratio of 2. To assess the importance
of the size-asymmetry effects, we collated our predictions with those of
HNC/MSA in the RPM, and also with the corresponding to the classical
Poisson-Boltzmann approaches: URMGC and MGC, to exhibit the notable
quantitative and qualitative discrepances existing between punctual and
non-punctual EDL theories. The main conclusion of this paper is that
size-asymmetry is an essential improvement to consider in studies of the
SEDL since heightens ionic size correlation effects (e.g.
nonmonotonicity, charge inversion and charge reversal) already seen in the
RPM case and, consequently, unveils new and interesting phenomenology absent
in size-symmetric systems. As a conspicuous example of these novel effects,
it is evidenced that counterions do not always dominate in the PM-SEDL, as
it is suggested by the PB point of view, and the wherefores of this fact are
traced back to the detailed characteristics of the EDL structure. In
particular, the contrasting high-$\sigma _{0}$ behavior displayed by the PB
and HNC/MSA potential-charge relationships, i.e. the convergence, for 
$\sigma _{0}\rightarrow \infty$, between the URMGC and MGC 
$\psi_{0}(\sigma _{0})$ and $\psi _{IHP}(\sigma _{0})$ curves and the
non-merging between the HNC/MSA$_{PM}$ and HNC/MSA$_{RPM}$ ones, was
elucidated by the following rational: \textit{for a system with fixed
parameters, when }$\sigma _{0}$\textit{\ increases the RDFs of URMGC and
MGC, and all their integrals and associated EDL properties (e.g. }$\psi _{0}$
\textit{\ and }$\psi _{IHP}$\textit{), go similar, whereas 
for HNC/MSA}$_{PM} $\textit{\ and HNC/MSA}$_{RPM}$\textit{, even if their 
RDFs can
have the same second moment, in general, their corresponding shapes and
ensuing properties (}$\psi _{0}$\textit{\ and/or }$\psi _{IHP}$\textit{)
could be completely different}.

In forthcoming publications some other appealing phenomena occurring in
PM-SEDL, as overcharging \cite{Ji01} and the advent of anomalous differential
capacities \cite{Go03}, will be addressed by means of our integral equation
approach.

\begin{acknowledgments}
E. G.-T. thanks the support by CONACYT (NC0072) and PROMEP.
\end{acknowledgments}

\appendix

\section{Explicit MSA bulk direct correlation functions} 

The direct correlation functions of a PM bulk electrolyte in the mean 
spherical approximation have been obtained analytically by Blum 
\cite{Bl01,Bl02} and 
Hiroike \cite{Hi02} for the general case of $n$ species with arbitrary size 
and charge. The only restriction is the electroneutrality condition:

\begin{equation}
\sum_{i=1}^n \rho_i z_i = 0.
\end{equation}
 
The MSA direct correlation functions can be expressed as:

\begin{equation}
c_{ij}(s) = c^{hs}_{ij}(s) + c^{elec}_{ij}(s),
\label{c_total}
\end{equation}

\noindent where $c_{ij}^{hs}(s)$ is the hard-sphere contribution and 
$c_{ij}^{elec}(s)$ is the electrostatic part.

\subsection{Electrostatic contribution}

Following the Baxter method \cite{Ba01} Blum found that 
the excess properties of an electrolyte can be written in terms of the 
parameter $\Gamma$. When a binary mixture of hard charged spheres 
of diameters $R_1$ and $R_2$ and valences $z_1$ and $z_2$ 
is considered, this parameter 
can be obtained solving the trascendental algebraic equation 
(it is assumed for definiteness that $R_2 \geq R_1$):
 
\begin{equation}
\Gamma^2 = \pi \omega D(\Gamma),
\end{equation}

\noindent where 

\begin{equation}
D(\Gamma) = \rho_1 {X_1}^2(\Gamma) + \rho_2 {X_2}^2(\Gamma),
\end{equation}

\begin{equation}
X_1(\Gamma) = \frac{z_1}{1+\Gamma R_1} + \frac{R_1^2 \chi(\Gamma)}{1+\Gamma R_1},
\end{equation}

\begin{equation}
X_2(\Gamma) = \frac{z_2}{1+\Gamma R_2} + \frac{R_2^2 \chi(\Gamma)}{1+\Gamma R_2},
\end{equation}

\begin{equation}
\chi = \frac{ -c \left\{ \rho_1 R_1 z_1 (1 + \Gamma R_1)^{-1} + 
\rho_2 R_2 z_2 (1 + \Gamma R_2)^{-1} \right\} }{ 1 + c\left\{
\rho_1 R_1^3(1 + \Gamma R_1)^{-1} + \rho_2 R_2^3(1 + \Gamma R_2)^{-1}\right\} },
\end{equation}

\begin{equation}
\Upsilon = (1+\Gamma R_1)(1+\Gamma R_2),  
\end{equation}

\begin{equation}
c = \frac{\pi}{2} \left[ 1 - \frac{\pi}{6}(\rho_1 R_1^3 + \rho_2 R_2^3) \right]^{-1}
\end{equation}

\noindent and

\begin{equation}
\omega = \frac{e^2}{\epsilon k_B T },
\end{equation}

\noindent with $\rho_1$ and $\rho_2$ being the numerical densities of the two species.

If we define

\begin{equation}
\lambda = \frac{R_2-R_1}{2}, 
\end{equation}

\begin{equation}
R_{12} = \frac{R_1+R_2}{2}
\end{equation}

\noindent and

\begin{equation}
P_{ij} = \frac{z_i z_j e^2}{ \epsilon k_B T},
\end{equation}

\noindent then $c_{ij}^{elec}(s)$ ($i,j=1,2$) is given explicitly by

\begin{equation}
c_{12}^{elec}(s)  = \beta_0, 0 \leq s \leq \lambda,
\end{equation}

\begin{equation}
c_{12}^{elec}(s) = \alpha_0 s^{-1} + \alpha_1 + \alpha_2 s + \alpha_3 s^3, \lambda \leq s < R_{12},
\end{equation}

\begin{equation}
c_{12}^{elec}(s) = - P_{12} s^{-1}, s > R_{12},
\end{equation}

\noindent where

\begin{eqnarray}
\lefteqn{\beta_0 = 2\omega \left[ \left( \frac{\Gamma{R_1}^4 - 2{R_1}^3}{3(1+\Gamma R_1)} \right) \chi^2  + \right.}
{}
\nonumber \\
& & {} 
\left. \left( \frac{z_1(R_2-R_1)}{\Upsilon} \right)\chi - \left(\frac{z_1 z_2 \Gamma}{1+\Gamma R_2} \right) \right],
\end{eqnarray}

\begin{eqnarray}
\lefteqn{\alpha_0  =  \frac{\omega (R_2-R_1)^2 }{16 \Upsilon} \Bigg( \bigg[ 4(R_1^2+R_2^2)- 4 \Gamma^2 R_1^2 R_2^2}  
{}
\nonumber\\
& & {} 
- {(R_1-R_2)}^2 \Upsilon  \bigg] \chi^2 + \bigg[4 (z_1+z_2) \bigg] \chi + 4 z_1z_2 \Gamma^2 \Bigg), 
\nonumber \\
\end{eqnarray}

\begin{eqnarray}
\lefteqn{\alpha_1  = \frac{\omega}{\Upsilon}\Bigg( 
\bigg[z_1(R_2-R_1) + z_2(R_1-R_2) \bigg] \chi - 2 \Gamma z_1 z_2 
{}}
\nonumber\\
& & {}  - ( R_1 + R_2) z_1 z_2 \Gamma^2 + \bigg[ \frac{R_1^3}{3}(1 + \Gamma R_2)( \Gamma R_1 - 2 )  
\nonumber\\
& & {}
+ \frac{R_2^3}{3}(1 + \Gamma R_1)( \Gamma R_2 - 2 ) \bigg] \chi^2  \Bigg), 
\end{eqnarray}

\begin{eqnarray}
\lefteqn{\alpha_2 = \frac{\omega}{\Upsilon} \Bigg( \bigg[ z_1 + z_2 \bigg] \chi + \Gamma^2 z_1 z_2 + {} }  
\nonumber\\
& & {} 
\bigg[ R_1^2 + R_2^2 - \Gamma^2 R_1^2 R_2^2 - \frac{ (R_1 - R_2)^2 \Upsilon }{2}  
\bigg] \chi^2
\Bigg)
\nonumber\\
\end{eqnarray}

\noindent and

\begin{eqnarray}
\alpha_3 =  \frac{w \chi^2}{3}. 
\end{eqnarray}

\begin{equation}
c_{11}^{elec}(s) = i_0  + i_1 s + i_2 s^3,0 \leq s \leq R_1,
\end{equation}

\begin{equation}
c_{11}^{elec}(s) = - P_{11} s^{-1},s > R_1
\end{equation}

\noindent with

\begin{eqnarray}
\lefteqn{i_0 = \frac{\omega}{(1+\Gamma R_1)^2} \Bigg( - 2 \Gamma z_1^2 - 2 R_1 z_1^2 \Gamma^2 +
{} }
\nonumber \\
& & {}         
 \bigg[ 
\frac{2}{3} R_1^3 (\Gamma R_1 - 2)  (1+\Gamma R_1) \bigg] \chi^2
\Bigg), 
\end{eqnarray}

\begin{eqnarray}
\lefteqn{i_1  =  \frac{\omega}{(1+\Gamma R_1)^2} \Big( \left[ 
2 R_1^2 - (\Gamma R_1^2)^2 \right] \chi^2 } 
{}
\nonumber \\
& & {}       
+ \left[ 2 z_1 \right] \chi + \Gamma^2 z_1^2 \Big)
\end{eqnarray}

\noindent and

\begin{eqnarray}
i_2 =  \frac{w \chi^2}{3}. 
\end{eqnarray}

\noindent Finally,

\begin{equation}
c_{22}^{elec}(s) = j_0  + j_1 s + j_2 s^3, 0 \leq s \leq R_2,
\end{equation}

\begin{equation}
c_{22}^{elec}(s) = - P_{22} s^{-1}, s > R_2,
\end{equation}

\noindent where

\begin{eqnarray}
\lefteqn{j_0  =  \frac{\omega}{(1+\Gamma R_2)^2} \Bigg( 
- 2 \Gamma z_2^2 - 2 R_2 z_2^2 \Gamma^2 +
{} }
\nonumber \\
& & {}        
\bigg[ \frac{2}{3} R_2^3 (\Gamma R_2 - 2)  (1+\Gamma R_2) \bigg] \chi^2
\Bigg),
\nonumber \\
\end{eqnarray}

\begin{eqnarray}
\lefteqn{j_1  =  \frac{\omega}{(1+\Gamma R_2)^2} \bigg( 
\left[ 2 z_2 \right] \chi + \Gamma^2 z_2^2 +
{} }
\nonumber \\
& & {}        
\left[ 2 R_2^2 - (\Gamma R_2^2)^2 \right] \chi^2  \bigg)
\end{eqnarray}

\noindent and

\begin{eqnarray}
j_2  =  \frac{w \chi^2}{3}. 
\end{eqnarray}

\subsection{Hard sphere contribution}

The exact solution for the Percus-Yevick equation of a hard spheres mixture 
with $n$ species was analyticaly obtained by Lebowitz \cite{Le02}.
Following the notation of Lebowitz, we define for a binary mixture

\begin{equation}
\xi = \frac{\pi}{6}\left( \rho_1 R_1^3 + \rho_2 R_2^3 \right),
\end{equation}

\begin{eqnarray}
\lefteqn{a_1 =  \left( 1 - \xi \right)^{-3} \Bigg\{ 1+\xi +\xi^2 +
\frac{\pi}{6} R_1^3(\rho_1 + \rho_2)(1 + 2 \xi) 
{} }
\nonumber \\
& & {}
- \frac{\pi}{2}(R_2-R_1)^2 \rho_2 \bigg[ \frac{\pi}{6} R_2 \rho_1 R_1^3 +(R_1 + R_2)+
{} 
\nonumber\\
& & {}  
\frac{\pi}{6} R_1 R_2 ( \rho_1 R_1^2 + \rho_2 R_2^2)
\bigg] \Bigg\} + \frac{\pi}{2}R_1^3(1-\xi)^{-4} \Bigg\{ 
{}
\nonumber\\
& & {} ( \rho_1 + \rho_2 )[ 1 + \xi + \xi^2 ] - 
\frac{\pi}{2} \rho_1 \rho_2 (R_2 - R_1)^2 \bigg[ 
{}
\nonumber\\
& & {}
(R_1+R_2) + \frac{\pi}{6} R_1 R_2 ( \rho_1 R_1^2 + \rho_2 R_2^2 ) \bigg] \Bigg\},   
\end{eqnarray}

\begin{eqnarray}
\lefteqn{a_2  =  \left( 1 - \xi \right)^{-3} \Bigg\{ 1 + \xi +\xi^2 +
\frac{\pi}{6} R_2^3(\rho_1 + \rho_2)(1 + 2 \xi) 
{} } 
\nonumber\\
& & {}  
- \frac{\pi}{2}(R_2-R_1)^2 \rho_1 \bigg[ \frac{\pi}{6} R_1 \rho_2 R_2^3 +
(R_1 + R_2)+
{} 
\nonumber\\
& & {}
\frac{\pi}{6} R_1 R_2 ( \rho_1 R_1^2 + \rho_2 R_2^2)
\bigg] \Bigg\} + \frac{\pi}{2}R_2^3(1-\xi)^{-4} \Bigg\{ 
{} 
\nonumber\\
& & {}
( \rho_1 + \rho_2 )[ 1 + \xi + \xi^2 ]  - \frac{\pi}{2} \rho_1 \rho_2 (R_2 - R_1)^2 \bigg[ 
{}
\nonumber\\
& & {} (R_1+R_2) + \frac{\pi}{6} R_1 R_2 ( \rho_1 R_1^2 + \rho_2 R_2^2 ) \bigg] \Bigg\},   
\end{eqnarray}

\begin{equation}
d = \frac{\pi}{12} \left[ \rho_1 a_1 + \rho_2 a_2 \right],
\end{equation}

\begin{equation}
g_{11} (R_1) = \left\{ 1 + \frac{\xi}{2} + \frac{\pi}{4} \rho_2 R_2^2(R_1-R_2) \right\}
(1-\xi)^{-2}, 
\end{equation}

\begin{equation}
g_{22} (R_2) = \left\{ 1 + \frac{\xi}{2} + \frac{\pi}{4} \rho_1 R_1^2(R_2-R_1) \right\}
(1-\xi)^{-2}, 
\end{equation}

\begin{equation}
g_{12} (R_{12}) = \left[ R_2 g_{11}(R_1) + R_1 g_{22}(R_2) \right] 
\left( R_1 + R_2 \right)^{-1},
\end{equation}

\begin{equation}
b_1 = - \frac{\pi}{4} \left[ 4 \rho_1 R_1^2 {g_{11}}^2(R_1) +
\rho_2( R_1 + R_2 )^2 {g_{12}}^2(R_{12}) \right],
\end{equation}

\begin{equation}
b_2 = - \frac{\pi}{4} \left[ 4 \rho_2 R_2^2 {g_{22}}^2(R_2) +
\rho_1( R_1 + R_2 )^2 {g_{12}}^2(R_{12}) \right]
\end{equation}

\noindent and 

\begin{eqnarray}
\lefteqn{b = - \frac{\pi}{2} \bigg[ \rho_1 R_1 g_{11}(R_1) + 
{} }
\nonumber \\
& & {}
\rho_2 R_2 g_{22}(R_2) \bigg] (R_1 + R_2)g_{12}(R_{12}).
\end{eqnarray}

The $c_{ij}^{hs}(s)$ ($i,j=1,2$) are then given by

\begin{equation}
c_{ii}^{hs}(s) = - a_i - b_i s - d s^3, s \leq R_i, i=1,2, 
\end{equation}

\begin{equation}
c_{ii}^{hs}(s) = 0, s>R_i, i=1,2,
\end{equation}

\begin{equation}
c_{12}^{hs}(s) = c_{21}^{hs}(s) = - a_1,0 \leq s \leq \lambda,
\end{equation}

\begin{eqnarray}
\lefteqn{c_{12}^{hs}(s) = c_{21}^{hs}(s) = - a_1 - \gamma_0 s^{-1} 
{} }
\nonumber \\
& & {} 
- \gamma_1 - \gamma_2 s - d s^3,\lambda \leq s \leq R_{12},
\end{eqnarray}

\begin{equation}
c_{12}^{hs}(s) = c_{21}^{hs}(s) = 0, s > R_{12},
\end{equation}

\noindent where

\begin{equation}
\gamma_0 = \lambda^2 ( b - 3 d \lambda^2 ),
\end{equation}

\begin{equation}
\gamma_1 = 8 \lambda^3 d - 2 \lambda b
\end{equation}

\noindent and 

\begin{equation}
\gamma_2 = b - 6 \lambda^2 d.
\end{equation}

Finally, if we define the constants

\begin{equation}
A_1=i_0-a_1;~~A_2=i_1-b_1;~~A_3=i_2-d;
\end{equation}

\begin{equation}
B_1=j_0-a_2;~~B_2=j_1-b_2;~~B_3=j_2-d;
\end{equation}

\begin{equation}
E_1=\beta_0-a_1;~~F_1=\alpha_0-\gamma_0;
\end{equation}

\begin{equation}
F_2=\alpha_1-a_1-\gamma_1;~~F_3=\alpha_2-\gamma_2;~~F_4=\alpha_3-d;
\end{equation}

\noindent we can write the bulk correlation functions as

\begin{equation}
c_{ij}(s)= c_{ij}^{hs}(s) + c_{ij}^{elec}(s)=c_{ij}^{int}(s)+ c_{ij}^{ext}(s),
\end{equation}

where

\begin{equation}
c_{11}^{int}(s) = P_{11} s^{-1} + A_1 + A_2 s + A_3 s^3, 0 \leq s \leq R_1,
\end{equation}

\begin{equation}
c_{11}^{int}(s) = 0,s > R_1, 
\end{equation}

\begin{equation}
c_{22}^{int}(s) = P_{22} s^{-1} + B_1 + B_2 s + B_3 s^3, 0 \leq s \leq R_2, 
\end{equation}

\begin{equation}
c_{22}^{int}(s) = 0, s > R_2, 
\end{equation}

\begin{equation}
c_{12}^{int}(s) = c_{21}^{int}(s) = P_{12} s^{-1} + E_1 , 
0 \leq s \leq \lambda, 
\end{equation}

\begin{eqnarray}
\lefteqn{
c_{12}^{int}(s)=c_{21}^{int}(s)= P_{12} s^{-1} +
{} } 
\nonumber \\
& & {} 
F_1 s^{-1} + F_2 + F_3 s + F_4 s^3,
\lambda \leq s \leq R_{12}, 
{}
\end{eqnarray}

\begin{equation}
c_{12}^{int}(s)=c_{21}^{int}(s) = 0, s \geq R_{12},
\end{equation}

\begin{equation}
c_{11}^{ext}(s) = -P_{11}s^{-1}, s > 0,
\end{equation}

\begin{equation}
c_{22}^{ext}(s) =-P_{22}s^{-1},s > 0
\end{equation}

\noindent and

\begin{equation}
c_{12}^{ext}(s)=c_{21}^{ext}(s)=-P_{12}s^{-1}, s > 0.
\end{equation}

\section{Explicit form of the $I_i(r)$ and $H_i(r)$ terms in the HNC/MSA 
integral equations for the PM-SEDL}

Due to the radial symmetry of the system, the angular dependence 
($\theta,\phi$) of the equations (\ref{int1})-(\ref{int2}), (\ref{I1})-(\ref{I4}) and (\ref{H1})-(\ref{H4}) 
can be resolved since the angular integrals can be performed analytically.  

The final form of the $I_i(r)$ and $H_i(r)$ terms can be given if we define

\begin{equation}
J( x_f,l,r,t ) = \frac{1}{(l+2)rt} \left[ x_f^{l+2} -
\mid r - t \mid^{l+2} \right],
\end{equation}

\noindent and 

\begin{equation}
J^\prime(x_i,x_f,l,r,t) = 
\frac{1}{(l+2)rt} \left[ x_f^{l+2} - x_i^{l+2} \right].
\end{equation}

\noindent Thus, if 
   
\begin{equation}
\frac{D}{2} + \frac{R_1}{2} \leq r \leq \frac{D}{2} + \frac{3}{2} R_1,
\end{equation}

\noindent we have

\begin{equation}
I_1(r) = \int_{\frac{D}{2} + \frac{R_1}{2}}^{r+R_1} KI_1a(r,t) dt +
\int_{\frac{D}{2} + \frac{R_1}{2}}^{\infty} KI_1b(r,t) dt,
\end{equation}

\noindent and, if 

\begin{equation}
\frac{D}{2} + \frac{3}{2} R_1 < r < \infty,
\end{equation}

\noindent then

\begin{equation}
I_1(r) = \int_{r-R_1}^{r+R_1} KI_1a(r,t) dt +
\int_{\frac{D}{2} + \frac{R_1}{2}}^{\infty} KI_1b(r,t) dt,
\end{equation}

\noindent where

\begin{eqnarray}
\lefteqn{KI_1a(r,t) = 2 \pi g_1 (t) t^2 \bigg( P_{11} J(R_1,-1,r,t) + 
{} }
\nonumber \\
& & {} A_1 J(R_1,0,r,t) +  A_2 J (R_1,1,r,t) + 
{} 
\nonumber \\
& & {} A_3 J(R_1,3,r,t) \bigg)
\end{eqnarray}

\noindent and 

\begin{equation}
KI_1b(r,t) = - 2 \pi P_{11} g_1 (t) \frac{t}{r} \bigg( r+t - \mid r - t \mid
\bigg). 
\end{equation}

\noindent If

\begin{equation}
\frac{D}{2} + \frac{R_1}{2} \leq r < \infty,
\end{equation}

\noindent then 

\begin{equation}
I_2(r) = 4 \pi \left( \frac{ P_{11} R_1^2 }{2} + \frac{ A_1 R_1^3 }{3} + 
\frac{ A_2 R_1^4 }{4} + \frac{ A_3 R_1^6 }{6} \right).
\end{equation}

\noindent If

\begin{equation}
\frac{D}{2} + \frac{R_1}{2} \leq r \leq \frac{D}{2} + \frac{R_2}{2} + \lambda,
\end{equation}

\noindent then

\begin{eqnarray}
\lefteqn{I_3(r) = \int_{ \frac{D}{2} + \frac{R_2}{2} }^{r + \lambda} KI_3a(r,t) dt +
{} } 
\nonumber \\
& & {} \int_{r + \lambda}^{r+R_{12}} KI_3b(r,t) dt +
\int_{ \frac{D}{2} + \frac{R_2}{2} }^{\infty} KI_3c(r,t) dt, 
\nonumber \\
& & {}
\end{eqnarray}

\noindent if

\begin{equation}
\frac{D}{2} + \frac{R_2}{2} + \lambda<r<\frac{D}{2} + \frac{R_2}{2} + R_{12}, 
\end{equation}

\noindent then

\begin{eqnarray}
\lefteqn{I_3(r) = \int_{ \frac{D}{2} + \frac{R_2}{2} }^{r - \lambda} KI_3b(r,t) dt +
\int_{ r - \lambda}^{r + \lambda} KI_3a(r,t) dt {} }
\nonumber \\
& & {}
+ \int_{r + \lambda}^{r+R_{12}} KI_3b(r,t) dt +  
\int_{ \frac{D}{2} + \frac{R_2}{2} }^{\infty} KI_3c(r,t) dt, 
\nonumber \\
& & {}
\end{eqnarray}

\noindent and, if

\begin{equation}
\frac{D}{2} + \frac{R_2}{2} + R_{12} \leq r < \infty,
\end{equation}

\noindent then

\begin{eqnarray}
\lefteqn{I_3(r) = \int_{ r - R_{12} }^{r - \lambda} KI_3b(r,t) dt +
\int_{ r - \lambda}^{r + \lambda} KI_3a(r,t) dt  {} }
\nonumber \\
& & {}
+ \int_{r + \lambda}^{r+R_{12}} KI_3b(r,t) dt +  
\int_{ \frac{D}{2} + \frac{R_2}{2} }^{\infty} KI_3c(r,t) dt, 
\nonumber \\
& & {}
\end{eqnarray}

\noindent where 

\begin{eqnarray}
\lefteqn{KI_3a(r,t) = 2 \pi g_2 (t) t^2 \bigg( P_{12} J(\lambda,-1,r,t)  + 
{}}
\nonumber \\
& & {} E_1 J(\lambda,0,r,t) + (P_{12}+F_1)J^\prime( \lambda, R_{12},-1,r,t) + 
{}
\nonumber \\
& & {} F_2 J^\prime( \lambda, R_{12},0,r,t) + F_3 J^\prime (\lambda,R_{12},1,r,t) + 
{}
\nonumber \\
& & {} F_4 J^\prime(\lambda,R_{12},3,r,t)  \bigg),
\end{eqnarray}

\begin{eqnarray}
\lefteqn{KI_3b(r,t) = 2 \pi g_2 (t) t^2 \bigg(  F_2 J( R_{12},0,r,t) +
 {} }
\nonumber \\
& & {}  F_3 J( R_{12},1,r,t) + F_4 J(R_{12},3,r,t)+
\nonumber \\
& & {} (P_{12}+F_1)J( R_{12},-1,r,t)  \bigg)
\end{eqnarray}

\noindent and 

\begin{equation}
KI_3c(r,t) = - 2 \pi P_{12} g_2 (t) \frac{t}{r} \bigg( r+t - \mid r - t \mid
\bigg). 
\end{equation}

\noindent If

\begin{equation}
\frac{D}{2} + \frac{R_1}{2} \leq r < \infty,
\end{equation}

\noindent then

\begin{eqnarray}
\lefteqn{I_4(r) = 4 \pi \bigg( P_{12} \frac{\lambda^2}{2} + E_1 \frac{\lambda^3}{3}
+  {}}
\nonumber \\
& & {} ( F_1 + P_{12} )\frac{ (R_{12}^2 - \lambda^2) }{2} + 
F_2 \frac{ (R_{12}^3 - \lambda^3 )}{3} + 
\nonumber \\
& & {}
F_3 \frac{ (R_{12}^4 - \lambda^4) }{4} + 
F_4 \frac{ (R_{12}^6 - \lambda^6)}{6}\bigg).
\end{eqnarray}

\noindent If

\begin{equation}
\frac{D}{2}+\frac{R_2}{2} \leq r \leq\frac{D}{2}+ \frac{R_1}{2} + R_{12}, 
\end{equation}

\noindent then

\begin{eqnarray}
\lefteqn{H_1(r) = \int_{ \frac{D}{2} + \frac{R_1}{2} }^{r - \lambda} KH_1b(r,t) dt +
\int_{ r - \lambda}^{r + \lambda} KH_1a(r,t) dt  {}}
\nonumber \\
& & {}
+ \int_{r + \lambda}^{r+R_{12}} KH_1b(r,t) dt +  
\int_{ \frac{D}{2} + \frac{R_1}{2} }^{\infty} KH_1c(r,t) dt, 
\nonumber \\
& & {}
\end{eqnarray}

\noindent and, if

\begin{equation}
\frac{D}{2} + \frac{R_1}{2} + R_{12} < r <  \infty,
\end{equation}

\noindent then

\begin{eqnarray}
\lefteqn{H_1(r) = \int_{ r - R_{12}}^{ r - \lambda } KH_1b(r,t) dt +
\int_{ r - \lambda}^{r + \lambda} KH_1a(r,t) dt  {}}
\nonumber \\
& & {}
+ \int_{r + \lambda}^{r+R_{12}} KH_1b(r,t) dt +  
\int_{ \frac{D}{2} + \frac{R_1}{2} }^{\infty} KH_1c(r,t) dt, 
\nonumber \\
& & {}
\end{eqnarray}

\noindent where

\begin{eqnarray}
\lefteqn{KH_1a(r,t) = 2 \pi g_1 (t) t^2 \bigg( P_{12} J(\lambda,-1,r,t)  + 
 {}}
\nonumber \\
& & {} E_1 J(\lambda,0,r,t) + (P_{12}+F_1)J^\prime( \lambda, R_{12},-1,r,t) + 
 {}
\nonumber \\
& & {} F_2 J^\prime( \lambda, R_{12},0,r,t) + F_3 J^\prime (\lambda,R_{12},1,r,t) +
 {}
\nonumber \\
& & {} F_4 J^\prime(\lambda,R_{12},3,r,t)  \bigg),
\end{eqnarray}

\begin{eqnarray}
\lefteqn{KH_1b(r,t) = 2 \pi g_1 (t) t^2 \bigg( F_2 J( R_{12},0,r,t) +
{} }
\nonumber \\
& & {}
F_3 J( R_{12},1,r,t) + F_4 J(R_{12},3,r,t) + 
{}
\nonumber \\
& & {}  (P_{12}+F_1)J( R_{12},-1,r,t) \bigg)
\end{eqnarray}

\noindent and

\begin{equation}
KH_1c(r,t) = - 2 \pi P_{12} g_1(t) \frac{t}{r} \bigg( r+t - \mid r - t \mid
\bigg). 
\end{equation}

\noindent If

\begin{equation}
\frac{D}{2} + \frac{R_2}{2} \leq r < \infty,
\end{equation}

\noindent then

\begin{equation}
H_2(r) = I_4(r).
\end{equation}

\noindent If

\begin{equation}
\frac{D}{2} + \frac{R_2}{2} \leq r \leq \frac{D}{2} + \frac{3}{2}R_2,
\end{equation}

\noindent then

\begin{equation}
H_3(r) = \int_{\frac{D}{2} + \frac{R_2}{2}}^{r+R_2} KH_3a(r,t) dt +
\int_{\frac{D}{2} + \frac{R_2}{2}}^{\infty} KH_3b(r,t) dt,
\end{equation}

\noindent and, if

\begin{equation}
\frac{D}{2} + \frac{3}{2} R_2 < r < \infty,
\end{equation}

\noindent then

\begin{equation}
H_3(r) = \int_{r-R_2}^{r+R_2} KH_3a(r,t) dt +
\int_{\frac{D}{2} + \frac{R_2}{2}}^{\infty} KH_3b(r,t) dt,
\end{equation}

\noindent where

\begin{eqnarray}
\lefteqn{KH_3a(r,t) = 2 \pi g_2 (t) t^2 \bigg( P_{22} J(R_2,-1,r,t)  + 
 {}}
\nonumber \\
& & {} B_1 J(R_2,0,r,t) + B_2 J (R_2,1,r,t) + 
{}
\nonumber \\
& & {} B_3 J(R_2,3,r,t) \bigg)
\end{eqnarray}

\noindent and

\begin{equation}
KH_3b(r,t) = - 2 \pi P_{22} g_2 (t) \frac{t}{r} \bigg( r+t - \mid r - t \mid
\bigg). 
\end{equation}

\noindent Finally, if

\begin{equation}
\frac{D}{2} + \frac{R_2}{2} \leq r < \infty,
\end{equation}

\noindent then 

\begin{equation}
H_4(r) = 4 \pi \left( \frac{ P_{22} R_2^2 }{2} + \frac{ B_1 R_2^3 }{3} + 
\frac{ B_2 R_2^4 }{4} + \frac{ B_3 R_2^6 }{6} \right).
\end{equation}

\begin{figure}[htbp] 
\begin{center} 
\begin{minipage}{\linewidth} 
{\includegraphics[angle=0.0, width=\linewidth]{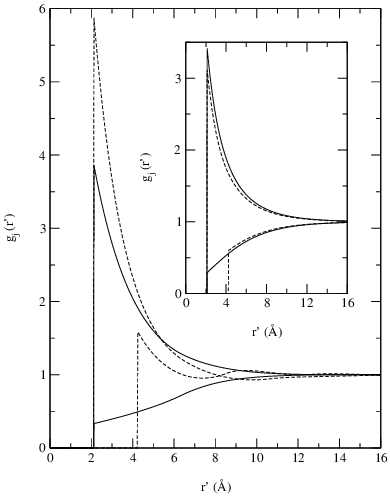}}
\caption{SEDL radial distribution functions for a 1:1, 1 M electrolyte with 
$R_{-}=4.25$~\AA\ and $R_{+}=8.5$~\AA\ next to a colloid of diameter $D=160$
\AA\ and surface charge density $\sigma _{0}=0.08356$ C/m$^{2}$. In the main
panel the solid and dashed lines stand for HNC/MSA$_{RPM}$ and HNC/MSA$_{PM}$,
respectively, whereas in the inset the solid and dashed lines correspond
to MGC and URMGC, respectively. The colloid-ion and ion-ion closest approach
distances used in each theory are the same as in Fig. \ref{fig_z0z1_c2}. 
Here (and in the rest of figures) the distance $r^{\prime }$ is measured 
from the macroparticle's surface.
} 
\label{new_gs_c2} 
\end{minipage} 
\end{center} 
\end{figure}
\begin{figure}[htbp] 
\begin{center} 
\begin{minipage}{\linewidth} 
{\includegraphics[angle=0.0, width=\linewidth]{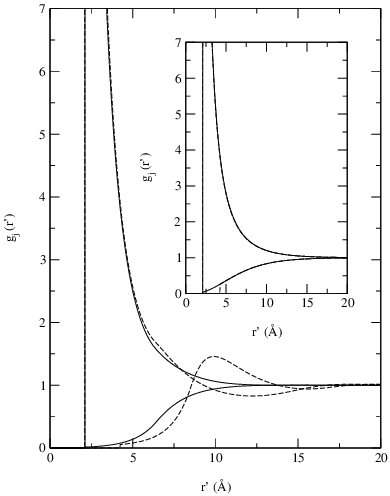}}
\caption{The same as in Fig. \ref{new_gs_c2} but for a surface charge 
density  $\sigma_{0}=0.3004$ C/m$^{2}$. 
} 
\label{new_gs_c2_3} 
\end{minipage} 
\end{center} 
\end{figure}
\begin{table}[htbp]
\caption{\label{t1} Parameters of the 2:2, 0.5  electrolytes used in 
the Monte Carlo and molecular dynamics simulations of the asymmetric SEDL. 
A macroion with diameter $D=10$ \AA\ and surface charge density 
$\sigma _{0}=0.407$ C/m$^{2}$ is employed in all the runs.
}
\begin{ruledtabular}
\begin{tabular}{cccccccccc}
Run & Method & $z_{1}$ & $z_{2}$ & $R_{1}$ (\AA) & $R_{2}$ (\AA) & $N_{1}$ & $N_{2}$  & \textit{L} \\
\hline
A & MC & $-2$ & $+2$ & 4.25 & 8.50  & 1308 & 1304 & 38.35 $R_1$ \\
B & DM & $-2$ & $+2$ & 4.25 & 8.50  & 2894 & 2890 & 50.00 $R_1$ \\
C & MC & $+2$ & $-2$ & 4.25 & 8.50  & 1304 & 1308 & 38.35 $R_1$ \\
D & DM & $+2$ & $-2$ & 4.25 & 8.50  & 2890 & 2894 & 50.00 $R_1$ \\
\end{tabular}
\end{ruledtabular}
\end{table}
\begin{table*}[htbp]
\caption{\label{t2} SEDL values of the radial distribution functions at 
$C_{-}\equiv(D+R_{-})/2$ and $C_{+}\equiv (D+R_{+})/2$\ for 1:1, 1 M and 
2:2, 0.5 M
size-asymmetric electrolytes around a colloid of diameter $D=160$ \AA\ and
variable $\sigma _{0}$, obtained from the HNC/MSA$_{PM}$, HNC/MSA$_{RPM}$,
URMGC\ and MGC theories. The ionic species have diameters $R_{-}=4.25$ \AA\
for the counterions and $R_{+}=8.5$ \AA\ for the coions. The colloid-ion and
ion-ion closest approach distances used in each theory are the same as in
Fig. \ref{fig_z0z1_c2}. The surface charge density $\sigma_0$ is in $C/m^2$.
}
\begin{ruledtabular}
\begin{tabular}{cccccccccc}
Electrolyte & $\sigma_0$ & $g^{\text{HNC/MSA}_{PM}}_-$ &  $g^{\text{HNC/MSA}_{PM}}_+$ & $g^{\text{HNC/MSA}_{RPM}}_-$ &  $g^{\text{HNC/MSA}_{RPM}}_+$  & $g^{URMGC}_-$ &  $g^{URMGC}_+$ & $g^{MGC}_-$ &  $g^{MGC}_+$ \\
\hline
1:1 & 0.08356 & 5.87  & 1.6  & 3.7  & 0.3  & 3.1   & 0.6 & 3.4  & 0.3  \\
1:1 & 0.3004  & 28.8 & 0.06 & 25.4 & 0.01 & 24.4  & 0.3 & 24.6 & 0.04  \\
2:2 & 0.08356 & 6.2  & 0.5  & 5.6   & 0.2  & 4.9  & 0.6 & 5.3  & 0.2   \\
2:2 & 0.3004  & 50.7 & 0.03 & 49.7 & 0.009 & 47.8 & 0.3 & 48.1& 0.02   \\
\end{tabular}
\end{ruledtabular}
\end{table*}

\begin{table*}[htbp]
\caption{\label{t3} SEDL mean electrostatic potentials at the macroion's 
surface ($\psi _{0}$) and at the inner Helmholtz plane ($\psi_{IHP}$) for 
the HNC/MSA$_{PM}$, HNC/MSA$_{RPM}$, URMGC and MGC theories. The parameters 
are the same as in Table \ref{t2}. The surface charge density $\sigma_0$ is 
in $C/m^2$ and the MEPs in $mV$.
}
\begin{ruledtabular}
\begin{tabular}{cccccccccc}
Electrolyte & $\sigma_0$ & $\psi^{\text{HNC/MSA}_{PM}}_{0}$ &  $\psi^{\text{HNC/MSA}_{PM}}_{IHP}$ & $\psi^{\text{HNC/MSA}_{RPM}}_{0}$ &  $\psi^{\text{HNC/MSA}_{RPM}}_{IHP}$ & $\psi^{URMGC}_{0}$ &  $\psi^{URMGC}_{IHP}$ & $\psi^{MGC}_{0}$ &  $\psi^{MGC}_{IHP}$\\
\hline
1:1 & 0.08356 & 33.4 & 8.5 & 47.8 & 22.9 & 54.1 & 29.2 & 56.4 & 31.5 \\
1:1 & 0.3004  &124.3 &34.8 &147.4 &58.0  &171.5  & 82.0 & 171.7 & 82.2\\
2:2 & 0.08356 & 26.7 & 1.8 & 36.2 & 11.3 & 45.3  & 20.4 & 46.3  & 21.4\\
2:2 & 0.3004  & 82.6 &-6.9 & 94.2 & 4.8  & 139.2 & 49.7 & 139.2 & 49.7\\
\end{tabular}
\end{ruledtabular}
\end{table*}


\begin{thebibliography}{99}
\bibitem{Tx01} In Ref. \cite{Kh01} the MSA\ expression for the potential 
difference
across a PM-SEDL was obtained, however, such formula has only been used in
the planar limit in Ref. \cite{Va02}.
\end{thebibliography}
\end{document}